\documentclass[floatfix,twocolumn,showpacs,preprintnumbers,amsmath,amssymb,pra,superscriptaddress,longbibliography]{revtex4-1}
\usepackage{color}
\usepackage[usenames,dvipsnames,svgnames,table]{xcolor}
\usepackage[colorlinks=true,linkcolor=blue,urlcolor=blue,citecolor=blue]{hyperref}
\usepackage{mathtools}
\usepackage{graphicx}
\usepackage{dcolumn}
\usepackage{array}
\usepackage{lipsum}
\usepackage{bm}
\usepackage{subfigure}
\usepackage{amssymb}
\usepackage{multirow}
\usepackage{tabularx}
\usepackage{amsmath}
\usepackage{braket}
\usepackage[capitalise]{cleveref}
\graphicspath{{plots/}}
\usepackage{lipsum}
\usepackage{comment}
\usepackage{mathrsfs}

\begin{document}
\title{
\textit{Ab initio} path integral Monte Carlo simulations of hydrogen snapshots\\ at warm dense matter conditions
}

\author{Maximilian B\"ohme}\email{m.boehme@hzdr.de}
\affiliation{Center for Advanced Systems Understanding (CASUS), D-02826 G\"orlitz, Germany}
\affiliation{Helmholtz-Zentrum Dresden-Rossendorf (HZDR), D-01328 Dresden, Germany}
\affiliation{Technische  Universit\"at  Dresden,  D-01062  Dresden,  Germany}

\author{Zhandos A. Moldabekov}
\affiliation{Center for Advanced Systems Understanding (CASUS), D-02826 G\"orlitz, Germany}
\affiliation{Helmholtz-Zentrum Dresden-Rossendorf (HZDR), D-01328 Dresden, Germany}


\author{Jan Vorberger}
\affiliation{Helmholtz-Zentrum Dresden-Rossendorf (HZDR), D-01328 Dresden, Germany}

\author{Tobias Dornheim}

\affiliation{Center for Advanced Systems Understanding (CASUS), D-02826 G\"orlitz, Germany}
\affiliation{Helmholtz-Zentrum Dresden-Rossendorf (HZDR), D-01328 Dresden, Germany}

\begin{abstract}
We combine ab initio path integral Monte Carlo (PIMC) simulations with fixed ion configurations from density functional theory molecular dynamics (DFT-MD) simulations to solve the electronic problem for hydrogen under warm dense matter conditions [M.~B\"ohme \emph{et al.}~Phys.~Rev.~Lett.~\textbf{129}, 066402]. The problem of path collapse due to the Coulomb attraction is avoided by utilising the \textit{pair approximation}, which is compared against the simpler Kelbg pair-potential. We find very favourable convergence behaviour towards the former. Since we do not impose any nodal restrictions, our PIMC simulations are afflicted with the notorious fermion sign problem, which we analyse in detail. While computationally demanding, our results constitute an exact benchmark for other methods and approximations within DFT. Our set-up gives us the unique capability to study important properties of warm dense hydrogen such as the electronic static density response and exchange--correlation (XC) kernel without any model assumptions, which will be very valuable for a variety of applications such as the interpretation of experiments and the development of new XC functionals.
\end{abstract}

\maketitle

\section{Introduction}
The quantum mechanical description of the hydrogen atom proposed by Erwin Schr\"odinger in 1926 \cite{Schroedinger} was one of the important milestones in the development of quantum mechanics. As the most abundant element in the universe, hydrogen plays a vital role ranging from technological applications \cite{Klinger2019,Moses_NIF} to the understanding of astrophysical objects \cite{Militzer_2008,saumon1}. 96 years after the solution to the hydrogen atom has been published, our theoretical understanding of this element still contains large gaps. It is of utmost importance to have this precise understanding for applications like energy generation from nuclear fusion \cite{Bethe1939}, which has the potential to provide an abundance of cheap and clean energy for the next millennia. Intriguingly, solid hydrogen was predicted to be a high temperature superconductor by Ashcroft \cite{AshcroftHydrogen68}. The observation of this transition has yet to be experimentally confirmed \cite{eremets2011conductive}. All of these remarkable properties make hydrogen a very interesting element to study in the high temperature and pressure regime.

However, despite almost 100 years of enormous research efforts, there still remain many important open questions about the many-body effects and bulk properties of hydrogen. The theoretical description of the simplest element in the periodic table has proven itself to be quite difficult. Even computationally and theoretically sophisticated schemes like density functional theory (DFT) are unable to accurately predict the liquid-liquid phase transition (LLPT) from an insulated molecular fluid to a conducting atomic liquid. 
In particular, DFT results for the LLPT strongly depend on the particular choice for the XC functional \cite{Pierleoni_PNAS_2016}, which has to be supplied as an empirical external input.
The atomic liquid metal-insulator transition is especially challenging to study using DFT since nuclear quantum effects influence the molecular bond of liquid hydrogen in the high pressure regime \cite{MoralesPRL2013}.

Of particular interest is the so-called warm dense matter (WDM) regime, where both the quantum coupling parameter $r_s=d/a_\textnormal{B}$ (with $d$ and $a_\textnormal{B}$ being the average interparticle distance and first Bohr radius) and degeneracy temperature $\Theta=k_\textnormal{B}T/E_\textnormal{F}$ (with $E_\textnormal{F}$ being the usual Fermi energy~\cite{quantum_theory,Ott2018}) are of the order of unity. Indeed, the accurate theoretical description of WDM constitutes a most formidable challenge~\cite{wdm_book,new_POP}, as it must take into account the complex interplay of a number of physical effects. Moreover, the development of accurate XC-functionals for thermal DFT simulations of WDM is still in its infancy~\cite{PhysRevLett.112.076403,groth_prl,PhysRevB.101.245141,PhysRevLett.120.076401} and constitutes an important bottleneck.

QMC methods, on the other hand, are in principle exact and have already been successfully deployed for the warm dense uniform electron gas (UEG) \cite{review,Malone_PRL_2016,dornheim_prl,groth_prl,dornheim_HEDP,dornheim_dynamic,dornheim_ML}. However, a significant pitfall of QMC methods for fermionic systems is the notorious fermionic sign problem \cite{LohPRB1990,troyer,dornheim_sign_problem}, which causes an exponential increase in the necessary Monte-Carlo steps with important parameters such as the system size to control the statistical error of any measured observable. 
In particular, Troyer and Wiese~\cite{troyer} have shown that the sign problem is NP-hard for some applications. 
A possible way to lift the fermion sign problem in PIMC simulations is to use restricted paths that do not cross any nodes of the density matrix~\cite{ceperley1991fermion,Brown_PRL_2013}. This restricted PIMC (RPIMC) method has already been successfully applied to low density hydrogen \cite{militzer2001path} and is used in a number of other applications such as the recently published first-principles equation of state (FPEOS) table by Militzer \emph{et al.}~\cite{MilitzerFPEOS}. One major disadvantage of the RPIMC method is that it requires precise knowledge of the nodal structure of the density matrix \cite{ceperley1991fermion}, which is, in general, not known and which is commonly approximated by the Slater determinant of single particle density matrices. This approximation has been shown to be inaccurate for the UEG for low temperature and high density \cite{Schoof2015PRL}.

To remedy this unsatisfactory situation, we have recently shown~\cite{Boehme2022} that it is indeed possible to 
carry out PIMC simulations of hydrogen over parts of the WDM regime without fixed nodes. Specifically we have utilised the unrestricted, direct PIMC method~\cite{cep,Dornheim_permutation_cycles,dornheim_sign_problem} for electrons in a static external potential given by ion snapshots taken from DFT-MD simulations. Throughout this work, we will refer to this approach as snapshot PIMC (snap-PIMC). In Ref.~\cite{Boehme2022}, we presented the first result for the static electronic density response of hydrogen, which enabled us to study the exchange correlation effects in this system. From these investigations, we were able to extract a static XC Kernel, which, in turn, was used in linear-response time-dependent DFT (LR-TD-DFT) to calculate the electronic dynamic structure factor of hydrogen for $r_s=2,4$. More specifically, we have observed that the commonly used adiabatic LDA (ALDA) breaks down at $r_s\geq4$ and $\Theta=1$. In fact, ALDA even performed worse than a pure mean-field calculation. 
We are convinced that the possibility to obtain the DSF based on an exact treatment of XC effects on the static level 
constitutes a promising route to improve the agreement between simulations and applications such as state-of-the-art X-ray Thomson scattering experiments~\cite{siegfried_review,kraus_xrts,ITCF}.

In this work, we give a detailed overview of the employed set-up for the PIMC simulations of hydrogen. As a first step, 
we have overcome the problem of path collapse due to the Coulomb attraction 
by utilizing the pair-approximation (PA) \cite{Militzer_HTDM_2016,cep} as well as the simpler Kelbg potential \cite{Filinov_PRE_2004}, which has been extensively used in PIMC simulations by Filinov and co-workers~\cite{Filinov2001,Filinov_2001,Filinov_CPP_2004}. 
While both approaches give the same result in the limit of a large number of imaginary-time propagators $P$~\cite{Filinov_PRE_2004}, we find a significantly improved convergence of the PA compared to Kelbg.
Since we do not impose any nodal restrictions on the paths, snap-PIMC suffers from the fermion sign problem. We show that the presence of the ions indeed makes the fermion sign problem more severe compared to the UEG in most cases. 
Still, PIMC simulations are feasible over substantial parts of the relevant parameter regime.
In addition, we investigate the impact of temperature and density on the real space density, 
which is the essential quantity governing the celebrated Hohenberg-Kohn theorems \cite{HohenbergKohn1964,martin_2004}. The theorems state that the ground-state electronic density of a system $n_0(\mathbf{r})$ uniquely determines its properties and even constructs a functional $E[n]$ with a global minimum at $n_0(\mathbf{r})$. This approach was extended to finite temperatures by Mermin~\cite{Mermin_1965}. Since we are able to compute the electronic density exactly without the empirical input of the XC-functional, 
we are in the unique position to rigorously benchmark corresponding DFT calculations. 

We are convinced that this work, as already demonstrated in Ref.~\cite{Boehme2022}, has the potential to study hydrogen on a true \textit{ab initio} level and will be highly useful for a gamut of applications, 
such as the development of accurate XC functionals for warm dense hydrogen 
and the 
interpretation of XRTS spectra from 
DT implosions~\cite{Poole2022}.

The paper is organised as follows. In \cref{sec:Theory}, the theoretical fundamentals are established including the system Hamiltonian (\ref{ssec:system}), a description of the PIMC method for hydrogen snapshots (\ref{ssec:PIMC}), and a detailed description on how to construct pair-interactions for attractive Coulomb potentials using the \textit{pair approximation} (PA) (\ref{ssec:pairapprox}). Furthermore, we give an overview of the Fermion sign problem (\ref{ssec:signProblem}) and the estimation of observables (\ref{ssec:estimators}), and show how to compute the density response in snap-PIMC (\ref{ssec:density_response}) by applying an external harmonic perturbation~\cite{moroni,moroni2,dornheim_pre,groth_jcp,Dornheim_PRL_2020}. In \cref{sec:results}, we present our simulation results, starting with an analysis of the convergence behaviour of snap-PIMC with  the number of imaginary time-slices $P$ in Sec.~(\ref{ssec:convergence}). Moreover, a study of the Fermion problem in snap-PIMC is presented in Sec.~(\cref{ssec:FSP_results}) and the effects of density and temperature on the real space electronic density are shown in Sec.~(\ref{ssec:rs}) and Sec.~(\ref{ssec:theta}). Finally, we compare our new, exact PIMC results with DFT simulations for the real space electronic density in Sec.~(\ref{ssec:DFT}). \cref{sec:summary} contains a concise summary of our method, and a discussion of the multitude of possible future applications.
\section{Theory}

\label{sec:Theory}

\subsection{System parameters and Hamiltonian}

\label{ssec:system}
The Hamiltonian of a hydrogen snapshot with the ionic positions $\{\mathbf{I}_0,\dots,\mathbf{I}_{N-1}\}$ is given by
\begin{equation}\label{eq:Hamiltonian}
\hat{H} = \underbrace{-\frac{1}{2} \sum_{l=1}^N \nabla_l^2}_{\hat K} + \underbrace{\hat{W} + \hat{V}_I\left(\{\mathbf{I}_0,\dots,\mathbf{I}_{N-1}\}\right)}_{\hat V}\ ,
\end{equation}
where the first two terms on the RHS correspond to the kinetic and interaction energy of the electrons, and the last term contains all ionic contributions.
Specifically, the interaction between both the electrons and nuclei is expressed by an Ewald sum over infinitely many respective periodic images. Following the notation by Fraser \textit{et al.}~\cite{Fraser_PRB_1996}, the resulting pair potential is written as
\begin{eqnarray}\label{eq:Ewald_pair}
\Psi(\mathbf{a},\mathbf{b})  &=& \frac{1}{L^3} \sum_{\mathbf{G}\neq\mathbf{0}}\left( \frac{e^{-\pi^2G^2/\kappa^2}}{\pi G^2}
e^{i2\pi \mathbf{G}\cdot(\mathbf{a}-\mathbf{b})}
\right)\\\nonumber & & - \frac{\pi}{\kappa^2 \Omega} + \sum_\mathbf{n} \frac{\textnormal{erfc}(\kappa|\mathbf{a}-\mathbf{b}+\mathbf{n}L|)}{|\mathbf{a}-\mathbf{b}+\mathbf{n}L|}\ ,
\end{eqnarray}
where $\mathbf{n}=(n_x,n_y,n_z)^T$ with $n_i\in\mathbb{Z}$, and $\mathbf{G}$ denote reciprocal lattice vectors without the factors of $2\pi$. Also note that Eq.~(\ref{eq:Ewald_pair}) is independent of the particular choice of the Ewald parameter $\kappa$, which can be exploited to accelerate the convergence of both (in principle infinite) sums. 
The electronic interaction is then given by
\begin{eqnarray}\label{eq:W}
\hat{W} = \sum_{l<k}^N \left[ \Psi(\hat{\mathbf{r}}_l,\hat{\mathbf{r}}_k)-\xi_\textnormal{M} \right]\ ,
\end{eqnarray}
where the Madelung constant $\xi_\textnormal{M}$ is being defined as
\begin{eqnarray}
\xi_\textnormal{M} = \lim_{\mathbf{a}\to\mathbf{b}}\left(\Psi(\mathbf{a},\mathbf{b})-\frac{1}{|\mathbf{a}-\mathbf{b}|}\right)
\end{eqnarray}
and takes into account the interaction by a point charge (electron or nucleus) with its own background and array of images.
Similarly, we define the external ionic potential term in Eq.~(\ref{eq:Hamiltonian}) as
\begin{eqnarray}
\hat{V}_I = &-& \sum_{l=1}^N\sum_{k=1}^N \left[\Psi(\hat{\mathbf{r}}_l,\mathbf{I}_k)-\xi_\textnormal{M} \right] \\\nonumber &+& \sum_{l<k}^N \left[\Psi(\mathbf{I}_l,\mathbf{I}_k)-\xi_\textnormal{M} \right] \ ,
\end{eqnarray}
where the bottom line solely contains interactions between the fixed ions and is thus simply given by a constant which does not affect the simulation.

\subsection{PIMC simulation of a hydrogen snapshot}
\label{ssec:PIMC}

\begin{figure}
    \centering
    \includegraphics[width=0.485\textwidth]{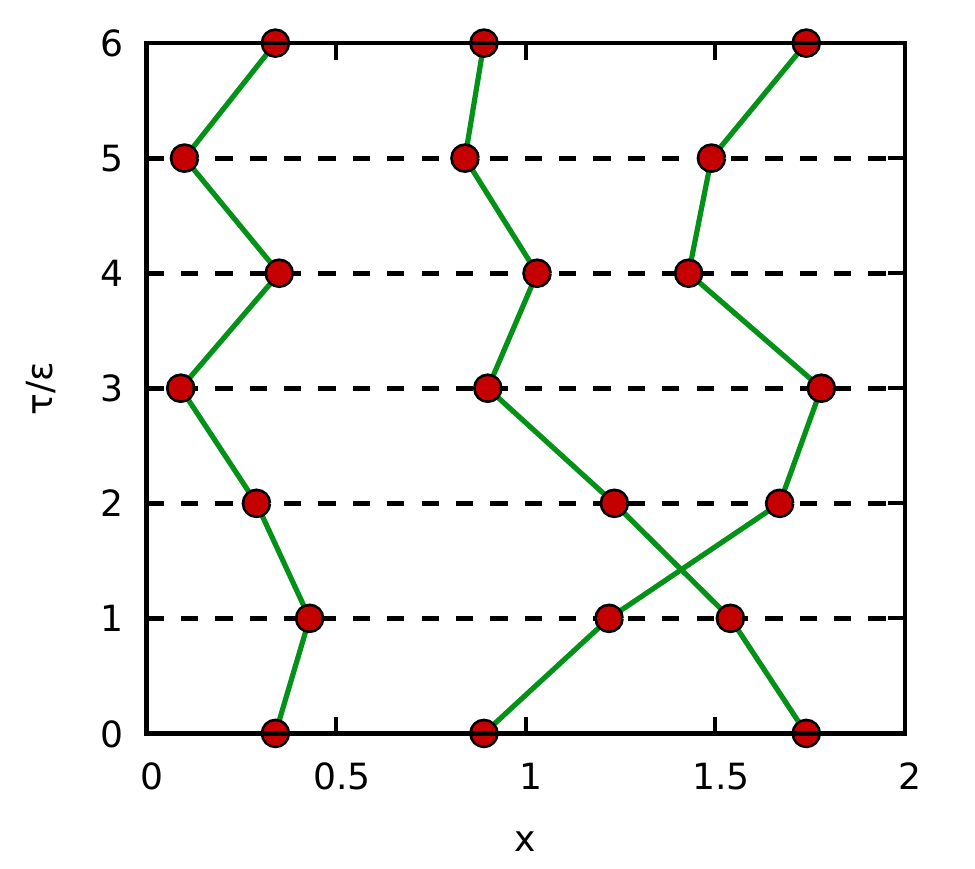}
    \caption{Illustration of imaginary-time paths in the $x$-$\tau$-plane for three particles. The red dots denote the so-called \textit{beads}, which depict the position of each particle at a particular imaginary-time step. The second and third path from the left illustrate a so-called exchange-cycle, where the trajectories of two particles are involved in a single path. Such \emph{permutation cycles}~\cite{Dornheim_permutation_cycles} are required to correctly take into account the indistinguishable nature of both bosons and fermions in the PIMC formalism.
    This exchange-cycle, ensures that PIMC abides the Pauli exclusion principle and also identical Bosons remain indistinguishable from one another \cite{boninsegni1}.
    }
    \label{fig:path-picture}
\end{figure}

We consider the canonical ensemble, meaning that the particle number $N$, volume $\Omega=L^3$, and inverse temperature $\beta=1/k_BT$ are fixed, where $k_B$ is the Boltzmann constant. Throughout this work, we use Hartree atomic units. Furthermore, we restrict ourselves to a fully spin-unpolarized system, i.e., $N^\uparrow = N^\downarrow = N/2$. The canonical partition function in coordinate space is then given by
\begin{align}\label{eq:Z}
Z_{\beta,N,V} &=& \frac{1}{N^\uparrow! N^\downarrow!} \sum_{\sigma^\uparrow\in S_{N^\uparrow}} \sum_{\sigma^\downarrow\in S_{N^\downarrow}} \textnormal{sgn}(\sigma^\uparrow,\sigma^\downarrow)\\\nonumber & & \times \int d\mathbf{R} \bra{\mathbf{R}} e^{-\beta\hat H} \ket{\hat{\pi}_{\sigma^\uparrow}\hat{\pi}_{\sigma^\downarrow}\mathbf{R}}\ ,
\end{align}
where $\mathbf{R}=(\mathbf{r}_1,\dots,\mathbf{r}_N)^T$ contains the coordinates of both spin-up and spin-down electrons. We again remark that we do not integrate over the ionic coordinates $\{\mathbf{I}_0,\dots,\mathbf{I}_{N-1}\}$, since they are fixed for the simulation of a hydrogen snapshot. In addition, $\hat{\pi}_{\sigma^\uparrow}$ ($\hat{\pi}_{\sigma^\downarrow}$) denotes the permutation operator corresponding to a particular element $\sigma^\uparrow$ ($\sigma^\downarrow$) from the permutation group $S_{N^\uparrow}$ ($S_{N^\downarrow}$), and the sign function $\textnormal{sgn}(\sigma^\uparrow,\sigma^\downarrow)$ is equal to positive (negative) unity for an even (odd) number of pair permutations~\cite{Dornheim_permutation_cycles}. Unfortunately, a straightforward evaluation of the matrix elements of the density operator
\begin{eqnarray}
\rho(\mathbf{R},\mathbf{R}';\beta) = \bra{\mathbf{R}} e^{-\beta\hat{H}} \ket{\mathbf{R}'}
\end{eqnarray}
is not possible as the kinetic and potential contributions to the full Hamiltonian [Eq.~(\ref{eq:Hamiltonian})] do not commute,
\begin{eqnarray}
e^{-\beta\hat{H}} \neq e^{-\beta\hat{K}} e^{-\beta\hat{V}}\ .
\end{eqnarray}
As a first step towards overcoming this obstacle, we employ the exact semi-group property of the density operator
\begin{eqnarray}\label{eq:group}
e^{-\beta\hat{H}} = \prod_{\alpha=0}^{P-1} e^{-\varepsilon\hat{H}}\ ,
\end{eqnarray}
with $P\in\mathbb{N}$ and the definition $\varepsilon=\beta/P$. Applying Eq.~(\ref{eq:group}) to Eq.~(\ref{eq:Z}) and inserting $P-1$ unity operators of the form
\begin{eqnarray}
\hat{1} = \int \textnormal{d}\mathbf{R}_\alpha\ \ket{\mathbf{R}_\alpha}\bra{\mathbf{R}_\alpha}
\end{eqnarray}
leads to the modified expression
\begin{eqnarray}\label{eq:Z_modified}
Z_{\beta,N,V} &=& \frac{1}{N^\uparrow! N^\downarrow!} \sum_{\sigma^\uparrow\in S_{N^\uparrow}} \sum_{\sigma^\downarrow\in S_{N^\downarrow}} \textnormal{sgn}(\sigma^\uparrow,\sigma^\downarrow)\\\nonumber & & \times \int d\mathbf{R}_0\dots d\mathbf{R}_{P-1}
\bra{\mathbf{R}_0}e^{-\varepsilon\hat H}\ket{\mathbf{R}_1}\\\nonumber & & \times \bra{\mathbf{R}_1}e^{-\varepsilon\hat H}\ket{\mathbf{R}_2} \dots 
\bra{\mathbf{R}_{P-1}} e^{-\varepsilon\hat H} \ket{\hat{\pi}_{\sigma^\uparrow}\hat{\pi}_{\sigma^\downarrow}\mathbf{R}_0}\ ,\label{eq:Z_modified_2} 
\end{eqnarray}
which is still exact. An illustration of the resulting paths in imaginary-time is given in \cref{fig:path-picture}. Here the red dots represent the so-called beads, which are the positions of the particles at a particular imaginary-time step $\tau$. Using the worm algorithm introduced in Ref.~\cite{boninsegni1}, one can now sample the canonical partition function by manipulating the beads using an efficient set of Monte-Carlo updates. The second and third path from the left show a so-called exchange cycle~\cite{Dornheim_permutation_cycles}. Such configurations ensure the adherence of the Pauli Exclusion Principle by taking into account the indistinguishable nature of the electrons.

Evidently, we have transformed the original expression for the canonical partition function into a high-dimensional integral over $P$ density matrices, each of which has to be evaluated at $P$ times the original temperature $T$. At this point, the task at hand is to find a suitable high-temperature approximation for $\rho(\mathbf{R},\mathbf{R}';\varepsilon)$ that becomes exact in the limit of large $P$. This can often be accomplished by employing the \emph{primitive factorization}
\begin{eqnarray}\label{eq:primitive}
e^{-\varepsilon\hat{H}} \approx e^{-\varepsilon\hat{V}}e^{-\varepsilon\hat{K}}\ ,
\end{eqnarray}
which is then justified by the well-known Trotter formula~\cite{trotter}
\begin{eqnarray}\label{eq:trotter}
\lim_{P\to\infty} \left( e^{-\varepsilon\hat{V}}e^{-\varepsilon\hat{K}} \right)^P = e^{-\beta(\hat{V}+\hat{K})}\ .
\end{eqnarray}
Unfortunately, Eq.~(\ref{eq:trotter}) only holds for potentials $\hat{V}$ that are bounded from below~\cite{kleinert2009path}. Obviously, this condition is violated by the snapshot Hamiltonian Eq.~(\ref{eq:Hamiltonian}) as the Coulomb attraction between an electron and an ion diverges towards negative infinity when the distance vanishes.

Let us define an effective interaction $\Phi(\mathbf{R},\mathbf{R}';\varepsilon)$ by the relation
\begin{eqnarray}
\rho(\mathbf{R},\mathbf{R}';\varepsilon) = e^{-\varepsilon\Phi(\mathbf{R},\mathbf{R}';\varepsilon)}\prod_{l=1}^N\rho_0(\mathbf{r}_l,\mathbf{r}_l';\varepsilon)\ ,
\end{eqnarray}
with the definition of the free particle density matrix
\begin{eqnarray}
\rho_0(\mathbf{r},\mathbf{r}';\varepsilon) = \frac{1}{\lambda_\varepsilon^{3/2}} e^{-\frac{\pi}{\lambda_\varepsilon^2}(\mathbf{r}'-\mathbf{r})^2}\ ,
\end{eqnarray}
and $\lambda_\varepsilon=\sqrt{2\pi\varepsilon}$ being the thermal wavelength associated with a single high-temperature factor. It is important to note that, while $\Phi(\mathbf{R},\mathbf{R}';\varepsilon)$ does have the dimension of an energy, it is off-diagonal in coordinate space and does not constitute an actual physical
interaction between a pair of particles. Within the primitive factorisation [Eq.~(\ref{eq:primitive})], it is simply given by
\begin{eqnarray}
\Phi_\textnormal{prim}(\mathbf{R},\mathbf{R}';\varepsilon) = \frac{1}{2}\left(
V(\mathbf{R}) + V(\mathbf{R}')
\right)\ ,
\end{eqnarray}
i.e., by the average of the potential energy between all electrons and ions evaluated on the two involved imaginary-time slices. Yet, the Coulomb divergence between electrons and ions is directly translated to $\Phi_\textnormal{prim}(\mathbf{R},\mathbf{R}';\varepsilon)$. Ultimately, this leads to a so-called \emph{path collapse}, as an electron can never be separated from an ion again once they are too close together.

A more suitable alternative is given by the \emph{pair approximation}~\cite{cep,Militzer_HTDM_2016} (\emph{PA}), which constructs the effective potential $\Phi(\mathbf{R},\mathbf{R}';\varepsilon)$ from a sum over pairs of particles,
\begin{eqnarray}\label{eq:pair}
e^{-\varepsilon\Phi_\textnormal{pair}(\mathbf{R},\mathbf{R}';\varepsilon)} = e^{-\varepsilon \sum_{l<k}^N \phi(\mathbf{r}_{lk},\mathbf{r}_{lk}';\varepsilon)}\ ,
\end{eqnarray}
where the effective pair potentials $\phi(\mathbf{r}_{lk},\mathbf{r}_{lk}';\varepsilon)$ are constructed from the nonideal parts of the (exact) pair density matrices $\rho(\mathbf{r}_{lk},\mathbf{r}_{lk}';\varepsilon)$ via the relation
\begin{eqnarray}\label{eq:pair_effective_potential}
\phi(\mathbf{r}_{lk},\mathbf{r}_{lk}';\varepsilon) = - \frac{1}{\varepsilon}\textnormal{log}\left(
\frac{\rho(\mathbf{r}_{lk},\mathbf{r}_{lk}';\varepsilon)}{\rho_0(\mathbf{r}_{lk},\mathbf{r}_{lk}';\varepsilon)}
\right)\ .
\end{eqnarray}
First and foremost, we note that Eq.~(\ref{eq:pair}) becomes exact in the limit of large $P$ as $P^{-4}$ \cite{Militzer_HTDM_2016,cep}, as three-body correlations and other higher order terms do not contribute when the temperature is high. Therefore, it constitutes a suitable scheme for the present study. Secondly, there is some freedom about how to exactly construct the total effective interaction as Eqs.~(\ref{eq:pair}) and (\ref{eq:primitive}) can be easily combined. Specifically, the only problematic term in the Hamiltonian [Eq.~(\ref{eq:Hamiltonian})] is given by the bare Coulomb attraction between an ion and the nearest image of an electron. This can be easily seen by considering the real-space part of the Ewald pair interaction Eq.~(\ref{eq:Ewald_pair}), which can be re-written as
\begin{eqnarray}\label{eq:decomposition}
\sum_\mathbf{n} \frac{\textnormal{erfc}(\kappa|\mathbf{a}-\mathbf{b}+\mathbf{n}L|)}{|\mathbf{a}-\mathbf{b}+\mathbf{n}L|} &=& \sum_\mathbf{n} \frac{1}{|\mathbf{a}-\mathbf{b}+\mathbf{n}L|} \\\nonumber & & - \sum_\mathbf{n} \frac{\textnormal{erf}(\kappa|\mathbf{a}-\mathbf{b}+\mathbf{n}L|)}{|\mathbf{a}-\mathbf{b}+\mathbf{n}L|}\ .
\end{eqnarray}
Evidently, the last term in Eq.~(\ref{eq:decomposition}) is always finite since it holds
\begin{eqnarray}
\lim_{x\to0}\frac{\textnormal{erf}(x\kappa)}{x} = \frac{2\kappa}{\sqrt{\pi}}\ .
\end{eqnarray}
We can thus decompose the total potential energy into the problematic nearest-image contribution $\hat{V}_\textnormal{NI}$ and the rest, $\hat{V}_\textnormal{R}=\hat{V}-\hat{V}_\textnormal{NI}$, with the definition
\begin{eqnarray}
\hat{V}_\textnormal{NI} = - \sum_{l=1}^N\sum_{k=1}^N \frac{1}{|\mathbf{I}_k-\hat{\mathbf{r}}_l|_\textnormal{NI}}\ ,
\end{eqnarray}
where $|\dots|_\textnormal{NI}$ denotes the absolute difference between the ionic position $\mathbf{I}_k$ and the nearest image of the electron at $\mathbf{r}_l$,
\begin{eqnarray}
|\mathbf{I}_k-\hat{\mathbf{r}}_l|_\textnormal{NI} = \textnormal{min}_{\mathbf{n}} |\mathbf{I}_k - \hat{\mathbf{r}}_l + \mathbf{n}L |\ .
\end{eqnarray}
The total effective potential $\Phi(\mathbf{R},\mathbf{R}';\varepsilon)$ can thus be constructed as
\begin{eqnarray}\label{eq:total_potential}
\Phi(\mathbf{R},\mathbf{R}';\varepsilon) = \Phi_\textnormal{NI}(\mathbf{R},\mathbf{R}';\varepsilon) + \Phi_\textnormal{R}(\mathbf{R},\mathbf{R}';\varepsilon)\ .
\end{eqnarray}
The last term in Eq.~(\ref{eq:total_potential}) can then simply be evaluated using the primitive approximation Eq.~(\ref{eq:primitive}) and the NI-term is constructed by evaluating Eq.~(\ref{eq:pair}) using as input the exact two-body Coulomb density matrix.

The practical details on the construction of the pair potential from the two-body Coulomb density matrix will be subject of the next section. 

While the evaluation of the \emph{PA} is possible in practice, it is worth considering if the exact numerical solution of the two-body Coulomb problem might be replaced by an approximate analytical expression that becomes exact in the limit of large $P$ sufficiently fast. Using first-order perturbation theory, Eq.~(\ref{eq:pair_effective_potential}) becomes~\cite{Filinov_PRE_2004}
\begin{eqnarray}\label{eq:perturbation}
\phi_\textnormal{NI}^0(\mathbf{r}_{lk},\mathbf{r}_{lk}';\varepsilon) = - \int_0^1 \textnormal{d}\nu\ \frac{\textnormal{erf}\left(\frac{|\nu\mathbf{r}_{lk}+(1-\nu)\mathbf{r}_{lk}'|}{2\lambda_{lk}\sqrt{\nu(1-\nu)}}\right)}{|\nu\mathbf{r}_{lk}+(1-\nu)\mathbf{r}_{lk}'|}\ ,
\end{eqnarray}
with the definition $\lambda_{lk}=\sqrt{\varepsilon/2\mu_{lk}}$ and the corresponding reduced mass $\mu_{lk}^{-1}=m_l^{-1}+m_k^{-1}$. Note that $\mathbf{r}_{lk}$ here automatically assumes the nearest-image convention, which is dropped for simplicity. While being considerably more simple than the full two-body density matrix, Eq.~(\ref{eq:perturbation}) still requires a numerical integration, which is too slow for PIMC simulations. A further simplification is given by considering the diagonal elements $\mathbf{r}=\mathbf{r}'$, which leads to the well-known Kelbg potential~\cite{kraeft2012quantum,bonitz_book,Filinov_PRE_2004},
\begin{eqnarray}\label{eq:Kelbg}
\phi_\textnormal{Kelbg}(\mathbf{r}_{lk};\varepsilon) = &-& \frac{1}{|\mathbf{r}_{lk}|}\left(
1 - e^{-\mathbf{r}_{lk}^2/\lambda_{lk}^2} \right. \\\nonumber &+& \left. \sqrt{\pi} \frac{|\mathbf{r}_{lk}|}{\lambda_{lk}}\left[
1 - \textnormal{erf}\left(\frac{|\mathbf{r}_{lk}|}{\lambda_{lk}}\right)
\right]
\right)\ ,
\end{eqnarray}
and the full effective potential is then simply obtained by averaging over the two involved imaginary-time slices,
\begin{eqnarray}\label{eq:Kelbg_action}
\Phi_\textnormal{NI,Kelbg}(\mathbf{R},\mathbf{R}';\varepsilon) &=& \frac{1}{2} \sum_{l=1}^N\sum_{k=1}^N \left[ 
\phi_\textnormal{Kelbg}(\mathbf{r}_{lk};\varepsilon)\right. \\\nonumber & & + \left. \phi_\textnormal{Kelbg}(\mathbf{r}_{lk}';\varepsilon)
\right]\ .
\end{eqnarray}

\begin{figure}\centering
\includegraphics[width=0.485\textwidth]{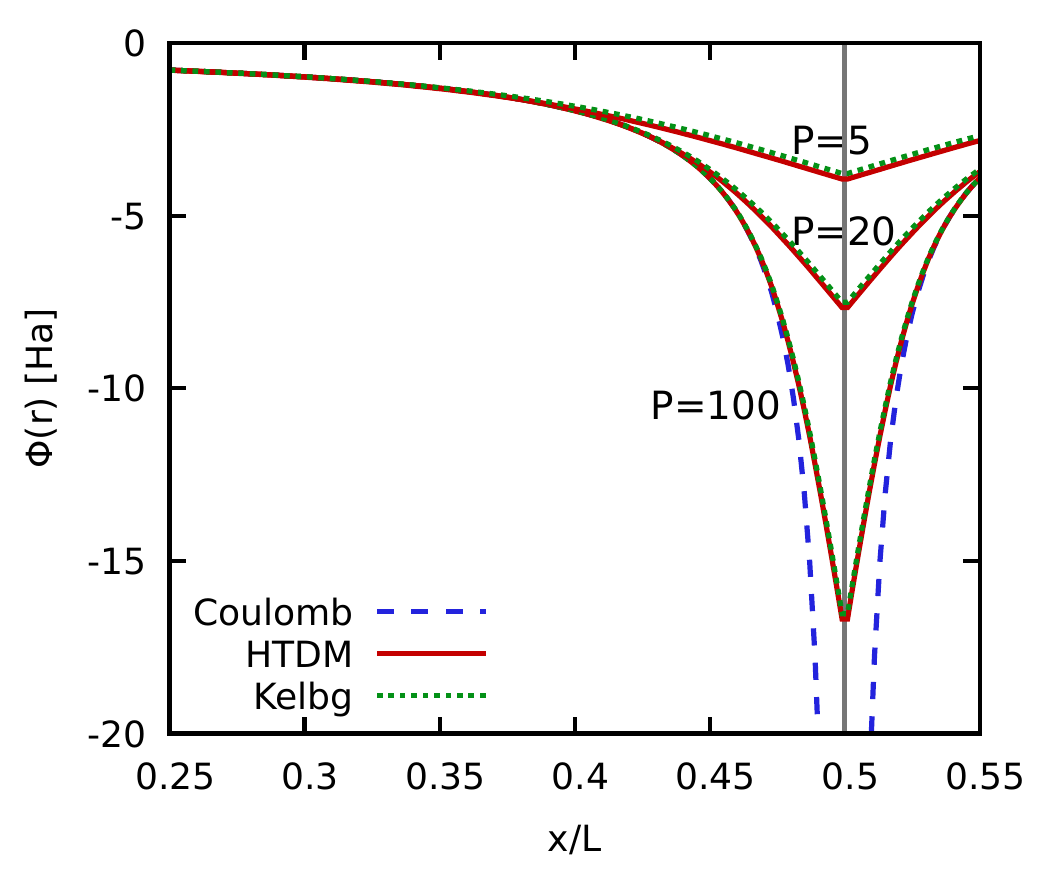}\\\vspace*{-0.93cm}
\includegraphics[width=0.485\textwidth]{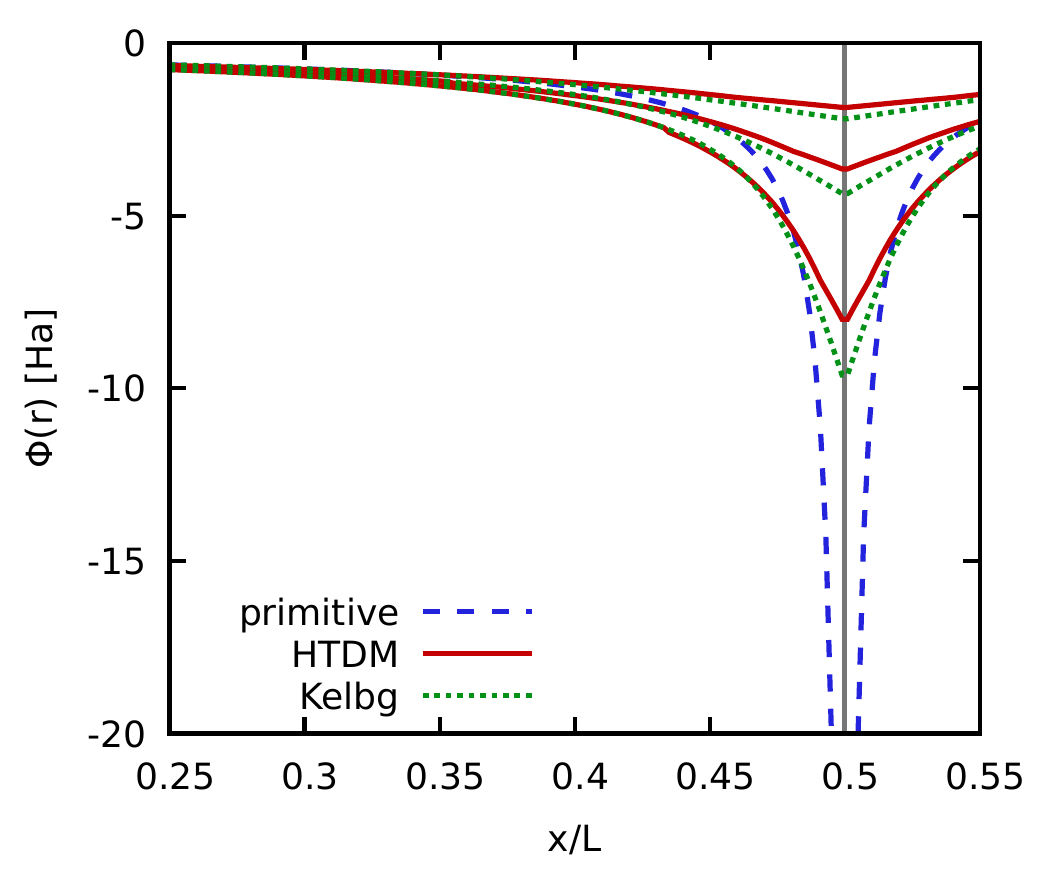}
\caption{\label{fig:info}
Comparison of various effective potentials between a test electron and an ion located at $\mathbf{I}=(L/2,0,0)^T$. Top panel: diagonal potential, $\mathbf{r}=\mathbf{r}'=(x,0,0)^T$. Bottom panel: off-diagonal potential, $\mathbf{r}=(x,0,0)^T$, $\mathbf{r}'(x,\lambda_\varepsilon,0)^T$.
}
\end{figure}

A practical demonstration of various effective potentials is shown in Fig.~\ref{fig:info} for $r_s=2$ and $\theta=1$. Specifically, the top panel shows results for the attractive potential between a single ion at $\mathbf{I}=(L/2,0,0)^T$ [where $L$ is the length of a simulation cell with $N=4$] and the nearest image of a test electron at $\mathbf{r}=\mathbf{r}'=(x,0,0)^T$. The dashed blue line corresponds to the bare Coulomb attraction (i.e., the primitive approximation), which diverges towards negative infinity for $x\to L/2$ as it is expected. The three solid red (dotted green) lines correspond to the effective potential due to the \emph{PA} (Kelbg potential) and have been obtained for $P=100$, $P=20$, and $P=5$ high-temperature factors; see the corresponding labels in the figure. First and foremost, we note that the high temperature density matrix (HTDM) and Kelbg curve are qualitatively very similar for all $P$. In particular, they attain a finite value around the position of the ion, which is of fundamental importance to avoid the phenomenon of path collapse in the PIMC simulations. Furthermore, they converge towards the bare Coulomb potential for large distances to the ion, which happens for smaller distances for larger numbers of high temperature factors $P$.

The bottom panel of Fig.~\ref{fig:info} shows the same information for an off-diagonal case, with the same $\mathbf{I}$ and $\mathbf{r}$, but $\mathbf{r}'=(x,\lambda_\varepsilon,0)^T$, i.e., a relative displacement of $\lambda_\varepsilon=\sqrt{2\pi\varepsilon}$ along the $y$-direction. Strictly speaking, this would lead to three distinct curves for the primitive approximation. Yet, as they can hardly be distinguished with the naked eye, we here restrict ourselves only to the case of $P=5$ as a reference. Furthermore, the off-diagonal nature does not affect the divergence of the bare Coulomb potential on the first time slice, which, in turn, leads to a diverging effective potential as well. Again, this issue is removed by both the Kelbg and HTDM-based potentials for all three values of $P$. Still, we note that the differences between the red and green curves are substantially larger than in the diagonal case, as the Kelbg potential by itself is not capable to intrinsically capture the off-diagonal nature of the action. This, in turn, implies a substantially slower convergence of PIMC expectation values with $P$, which is empirically verified in Sec.~\ref{ssec:convergence}. Finally, the primitive approximation only becomes accurate in the limit of very large distances between the electron and ion.

The final result for the partition function is given by
\begin{eqnarray} \nonumber
Z_{\beta,N,V} &=&  \int \textnormal{d}\mathbf{X}\ \textnormal{sgn}(\mathbf{X}) \left( \prod_{\alpha=0}^{P-1}\prod_{l=1}^N\rho_0(\mathbf{r}_{l,\alpha},\mathbf{r}_{l,\alpha+1};\varepsilon) \right)\\ & & \times \left( \prod_{\alpha=0}^{P-1} e^{-\varepsilon \Phi_\textnormal{R}(\mathbf{R}_\alpha)} e^{-\varepsilon\Phi_\textnormal{NI}(\mathbf{R}_\alpha,\mathbf{R}_{\alpha+1};\varepsilon)}\right)\ ,
\end{eqnarray}
where the integration over the multi-variable $\mathbf{X}=(\mathbf{R}_0,\dots,\mathbf{R}_{P-1})^T$ also contains the sum over all possible permutations of particle coordinates of the same spin.


\subsection{Constructing pair-potentials from the two-body Coulomb density matrix}
\label{ssec:pairapprox}
As mentioned in the previous section, we construct the pair-potential for the \emph{PA} using an analytical expression for the two-body Coulomb density matrix. The density matrix for a two-body system with Coulomb interaction is given by

\begin{widetext}
\begin{equation}
\label{eq:HTDM_vanilla}
\rho(\mathbf{r}, \mathbf{r}';\beta) = \sum_{l=0}^{\infty} \frac{ P_{l}(cos(\theta))}{4\pi} \underbrace{\sum_{n=l+1}^{\infty} \mathrm{e}^{-\beta E_{n}} R_{n}^{l}(r) R_{n}^{l}\left(r^{\prime}\right)+\int_{0}^{\infty} \mathrm{d} k \mathrm{e}^{-\beta \kappa k^{2}} F_{k}^{l}(r) F_{k}^{l}\left(r^{\prime}\right)}_{=:\rho_l(\mathbf{r}, \mathbf{r}';\beta)},
\end{equation}
\end{widetext}
with the distances in units of the Bohr radius $a_B$, the energies in Rydbergs and the dimensionless parameters as proposed in Ref.~\cite{Pollock_HTDM_1988}.

\begin{align}
    \kappa &:= \frac{\hbar^2}{2 \mu^2} e^2 a_B, \\
    Z &:= \frac{Z_1 Z_2}{\kappa}~.
\end{align}

Here $\mu^{-1} = (m_1^{-1} + m_2^{-1})$ refers to the reduced mass and $Z_i$ refers to the charge of the individual particle. $R_n^l(r)$ refers to the radial part of the eigenfunctions for the bound states in an attractive Coulomb potential and $F_n^l(r)$ refers to the radial Coulomb scattering eigenfunctions \cite{abramowitz+stegun}.
\cref{eq:HTDM_vanilla} is hard to converge sufficiently. The reason for that is the sum over the $l$-channel integrals containing the Coulomb scattering functions, which are hard to evaluate. 
To overcome this obstacle we use the result of Pollock \cite{Pollock_HTDM_1988}, who obtained an expression for the two body Coulomb density matrix using the results from Ref.~\cite{Hostler1963}
\begin{widetext}

\begin{equation}
\label{eq:HTDM_pollock}
\rho^{[3]}\left(\boldsymbol{r}, \boldsymbol{r}^{\prime} ; \beta\right)=\frac{1}{\pi} \frac{\partial}{\partial \eta} \rho_{l=0}^{[3]}\left(\frac{\xi+\sqrt{\xi^{2}-\eta}}{2}, \frac{\xi-\sqrt{\xi^{2}-\eta}}{2} ; \beta\right),
\end{equation} 
\end{widetext}

where $\xi = r + r'$ and $\eta = (r+r')^2 - |\mathbf{r} - \mathbf{r}'|^2$.
The resulting expression is much easier to converge than \cref{eq:HTDM_vanilla}, since it only results in a single evaluation of the integral over the Coulomb scattering functions. However, evaluating the above expression multiple times at each Monte-Carlo step would also be computationally very costly due to the numerical integration. A solution to this problem was proposed by Ceperley in Ref.~\cite{Ceperley1991} by constructing a lookup table of the resulting pair-action as a combination of interpolations and polynomial fits. We found the simplified polynomial ansatz from Militzer \cite{Militzer_HTDM_2016} easier to evaluate and sufficiently accurate to use in the snapshot PIMC simulations.
We note here that the estimator for the total energy also requires the $\beta$-derivative of \cref{eq:HTDM_pollock}. The resulting expression thus requires another lookup-table, which can again be accurately captured by the ansatz in Ref.~ \cite{Militzer_HTDM_2016}.
A distinct advantage for the usage of the Coulomb density matrix over the simplified Kelbg pair-potential is the significantly faster convergence with the amount of time-slices, which results in a much faster runtime and therefore faster convergence of the estimators. The convergence speed as a function of the number of time-slices is further investigated in \cref{ssec:convergence}. In this section we always draw the comparison against the Kelbg potential, which is analytically exact for a sufficiently large number of imaginary time steps. Any inaccuracy of Militzers ansatz would result in a deviation of the converged energy or induced density. Since both Kelbg and PA converge to the same value as a function of $P$ we determined the accuracy of this ansatz to be sufficient.

\subsection{The Fermion sign problem}
\label{ssec:signProblem}
A major obstacle of fermionic QMC is the notorious sign problem. It is a consequence of the anti-symmetry of any many-body fermionic wave-function under particle exchange. 
We start with the basic idea behind Metropolis Monte-Carlo sampling \cite{metropolis}, by drawing samples of configurations $\mathbf{X}$ from a probability distribution $W(\mathbf{X})$, with an unknown normalisation. In the case of PIMC, we wish to sample the canonical partition function Z, see \cref{eq:Z},
\begin{align}
    Z &= \int \, d\mathbf{X} \, W(\mathbf{X}),
\end{align}
where we identify the imaginary time-paths $\mathbf{R}=(\mathbf{R}_1, \dots, \mathbf{R}_{P-1})^T$ with the configuration of the system. Each vector $\mathbf{R}_i=(\mathbf{r}_{i,1}, \dots , \mathbf{r}_{i,N})$ contains all the particle positions at the imaginary time-slice $i$. One now can use an implementation of the Metropolis algorithm~\cite{metropolis} such as the schemes introduced in Refs.~\cite{boninsegni1,boninsegni2,Dornheim_PRB_nk_2021} in order to sample the configuration space of the system. The expectation value of any observable is then simply given by
\begin{equation}
\langle \hat{O} \rangle = \frac{1}{Z} \int d\mathbf{X} \, O(\mathbf{X}) \, W(\mathbf{X}).    
\end{equation}

However, in the case of fermionic systems one has to consider the antisymmetrisation of the wavefunction due to the Pauli exclusion principle. This has already been included in \cref{eq:Z} and leads to the issue that $W(\mathbf{X})$ can include negative weights, thus its interpretation as a probability density is not possible anymore. The Metropolis algorithm requires a probability distribution and therefore a straight-forward application of this algorithm is not possible. We resort to generate configurations according to the modified probability density

\begin{equation} \label{eq:distr_mod}
    P'(\mathbf{X}) = \frac{|W(\mathbf{X})|}{\int d\mathbf{X} \, |W(\mathbf{X})|} = \frac{W'(\mathbf{X})}{Z'},
\end{equation}
which is identical to the probability density of the corresponding bosonic system.
The fermionic expectation values therefore are now of the following form  
\begin{equation}
    \langle \hat{O} \rangle_f = \frac{\int d\mathbf{X} \, \textnormal{sgn}(W(\mathbf{X})) W'(\mathbf{X}) \hat{O}(\mathbf{X})}{\int d\mathbf{X} \, \textnormal{sgn}(W(\mathbf{X})) W'(\mathbf{X})},
\end{equation}
with $\textnormal{sgn}(\mathbf{X})=W(\mathbf{X}) / |W(\mathbf{X})| \equiv S(\mathbf{X})$ as the so called sign of the configuration. With this definition we are able to write the fermionic expectation values in the compact form

\begin{equation}
    \langle \hat{O} \rangle_f = \frac{\langle \hat{O} \hat{S}\rangle'}{\langle\hat{S}\rangle'},
\end{equation}
where the dashed expectation values denote that the modified probability density \cref{eq:distr_mod} is used. However, due to the sign terms it is now possible that cancellations occur during the Monte-Carlo sampling procedure. These cancellations affect the statistical uncertainty of any observable to a large extent. The average sign $\langle \hat{S} \rangle'$ therefore 
decisively determines the feasibility of a fermionic PIMC simulation for a given system.
It is easy to see that $\langle \hat S \rangle$ vanishes both for low temperatures and large systems as
\begin{equation}
\label{eq:sign_avg}
    \langle \hat S \rangle = e^{-\beta N (f - f') },
\end{equation}

where f and f' denote the free energy density of the fermionic and the bosonic system, respectively. The above relation has profound implications on the convergence of the resulting Monte-Carlo estimators

\begin{equation}
    \frac{\Delta O}{O} \sim \frac{1}{S \sqrt{M}} \sim \frac{e^{\beta N (f-f')}}{\sqrt{M}} .
\end{equation}
Thus, the statistical uncertainty of any observable will hit an exponential wall, when increasing the system size $N$  or the inverse temperature $\beta$. For a more detailed technical discussion in the case of different systems we refer the reader to Ref.~\cite{dornheim_sign_problem}. The results of the average sign in the case of snap-PIMC simulations of hydrogen are shown in \cref{ssec:FSP_results}.

\subsection{Estimation of observables}
\label{ssec:estimators}
The thermodynamic PIMC estimator for the total energy $E$ can be derived from the partition function via the relation
\begin{eqnarray}\label{eq:TD}
E = - \frac{1}{Z_{\beta,N,V}} \frac{\partial Z_{\beta,N,V}}{\partial\beta}\ .
\end{eqnarray}
In particular, the primitive approximation leads to the familiar expression
\begin{align}
\label{eq:Energy_primitive_approximation}
E_\textnormal{prim}(\mathbf{X}) &= \frac{3NP}{2\beta} - \frac{mP}{2\hbar^2\beta^2} \sum_{\alpha=0}^{P-1}\sum_{l=1}^N(\mathbf{r}_{l,\alpha}-\mathbf{r}_{l,\alpha+1})^2 \\\nonumber 
& + \frac{1}{P} \sum_{\alpha=0}^{P-1} V(\mathbf{R}_\alpha)\ ,
\end{align}
where the top and bottom line of the RHS. correspond to the kinetic and potential contributions, respectively. Let us next consider the modification of Eq.~(\ref{eq:Energy_primitive_approximation}) when we instead use the action 
given by the Kelbg potential [Eq.~(\ref{eq:Kelbg_action})]. Specifically, the Kelbg potential itself is diagonal in $\mathbf{R}$ and may thus simply be included into the total potential energy on a particular imaginary time slice $V(\mathbf{R})$. Yet, in contrast to the bare Coulomb (or Ewald) pair interaction, it does have an explicit dependence on the temperature, which, in turn, leads to an additional term upon evaluation of Eq.~(\ref{eq:TD}),
\begin{eqnarray}
E_\textnormal{Kelbg}(\mathbf{X}) = E_\textnormal{prim}(\mathbf{X}) +\frac{\beta}{P} \sum_{\alpha=0}^{P-1}\sum_{l=1}^N\Gamma(\mathbf{r}_{l,\alpha})\ ,
\end{eqnarray}
with the definition
\begin{eqnarray}
\Gamma(\mathbf{r}_{l,\alpha}) = - \sum_{k=1}^N\frac{\partial\phi_\textnormal{Kelbg}(\mathbf{r}_{l,\alpha}-\mathbf{I}_k;\varepsilon)}{\partial\beta}\ .
\end{eqnarray}
Obtaining the actual $\beta$-derivative of the Kelbg potential [Eq.~(\ref{eq:Kelbg})] is straightforward, and we find
\begin{eqnarray}
\frac{\partial\phi_\textnormal{Kelbg}(\mathbf{r}_{lk};\varepsilon)}{\partial\beta} = - \frac{\sqrt{\pi}}{2\lambda_{lk}\beta}\left\{
\textnormal{erf}\left(\frac{|\mathbf{r}_{lk}|}{\lambda_{lk}}\right)  - 1
\right\}\ .
\end{eqnarray}

In the case of the \emph{PA} we find a similar relation. The difference is that the last term in \cref{eq:Energy_primitive_approximation} now reads
\begin{align}
E_\textnormal{pair}(\mathbf{X}) = E_\textnormal{prim}(\mathbf{X}) + \sum_{\alpha=0}^{P-1} \frac{\partial u_\textnormal{pair}(\mathbf{R}_\alpha)}{{\partial \beta}},
\end{align}

with $u_\textnormal{pair}$ being the so-called pair action given by

\begin{equation}
u_\textnormal{pair} = \textnormal{log} \left( \frac{\rho_\textnormal{free}(\mathbf{r}, \mathbf{r}';\beta)}{\rho_\textnormal{C}(\mathbf{r}, \mathbf{r}';\beta)} \right).
\end{equation}
One therefore has to obtain the derivative of the two-body Coulomb density matrix given by \cref{eq:HTDM_pollock}. In order to improve performance, we used the method from Ref.~\cite{Militzer_HTDM_2016} and compute a lookup table of the corresponding derivative before running our simulation in addition to the lookup table of the HTDM.

\subsection{PIMC approach to the density response}

\label{ssec:density_response}

For uniform systems like the UEG or a hydrogen plasma, the entire wave-vector dependence of the static linear density response function $\chi(\mathbf{q})$ can be obtained from a single simulation of the unperturbed system by utilising the imaginary-time version of the well-known fluctuation dissipation theorem~\cite{Dornheim_JCP_ITCF_2021},
\begin{eqnarray}\label{eq:static_chi}
\chi^{(1)}(\mathbf{q}) = -n\int_0^\beta \textnormal{d}\tau\ F(\mathbf{q},\tau) \quad ,
\end{eqnarray}
where $F(\mathbf{q},\tau)$ denotes the usual imaginary-time version of the intermediate scattering function; see, e.g., Ref.~\cite{Dornheim_JCP_ITCF_2021} for details. Yet, Eq.~(\ref{eq:static_chi}) does not hold for a hydrogen snapshot as defined by the Hamiltonian from \cref{eq:Hamiltonian}, which is inhomogeneous even in the unperturbed case.

Therefore, we use the same methodology as already employed in Ref.~\cite{DornheimPRR2021} by modifying the snapshot Hamiltonian \cref{eq:Hamiltonian} to include a harmonic perturbation 
\begin{equation}
    \hat{H}_{\mathbf{q},A} = \hat{H}_{\textnormal{SNAP}} + 2A\sum_{i=1}^{N} \textnormal{cos}(\mathbf{q}\mathbf{r}_i),
\end{equation}
where A is the perturbation strength and $\mathbf{q} = (n_x,n_y,n_z)^T  2\pi/L$ the perturbation wave-vector. Using this modified Hamiltonian, we compute the density in reciprocal space,
\begin{equation}\label{eq:rho}
\braket{\hat\rho_\mathbf{k}}_{q,A} = \frac{1}{V} \left< \sum_{l=1}^N e^{-i\mathbf{k}\cdot\hat{\mathbf{r}}_l} \right>_{q,A} \ , 
\end{equation}
and the induced density is given by 
\begin{eqnarray}\label{eq:induced}
\Delta\rho_{\mathbb{q},A} \equiv \braket{\hat\rho_\mathbf{k}}_{q,A} - \braket{\hat\rho_\mathbf{k}}_{q,0}\ .
\end{eqnarray}
In the limit of small $A$, Eq.~(\ref{eq:induced}) can be expanded as
\begin{equation}
\label{eq:rho_expansion}
\braket{\hat{\rho}_\mathbf{q}}_{\mathbf{q},A} = \chi^{(1)}(\mathbf{q})A + \mathcal{O}(A^2),
\end{equation}
with $\chi(\mathbf{q})$ being linear response coefficient of interest.

\section{Results}
\label{sec:results}
\subsection{Convergence properties}
\label{ssec:convergence}
\begin{figure*}
\centering\includegraphics[width=0.45\textwidth]{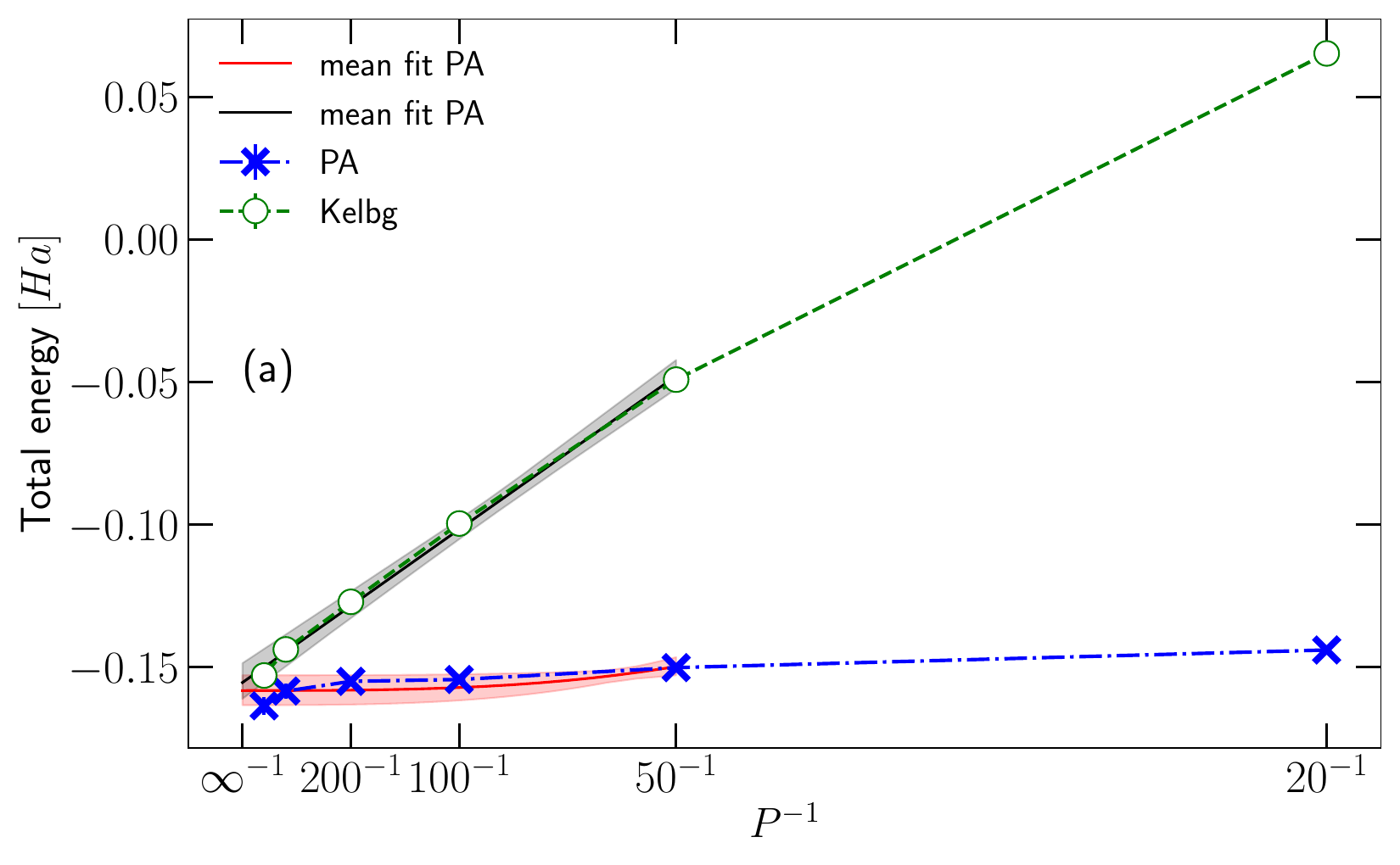}
\centering\includegraphics[width=0.45\textwidth]{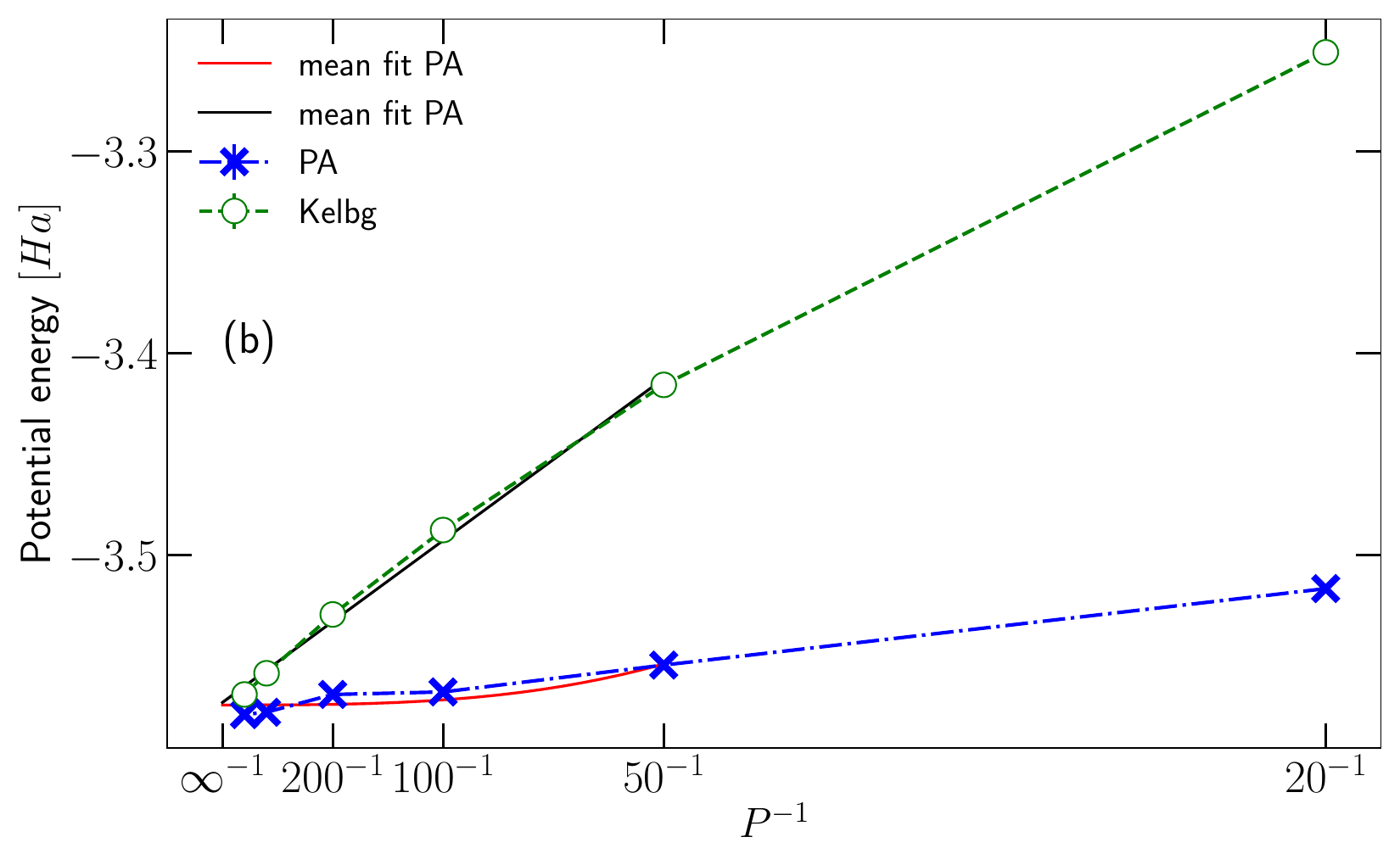}
\centering\includegraphics[width=0.45\textwidth]{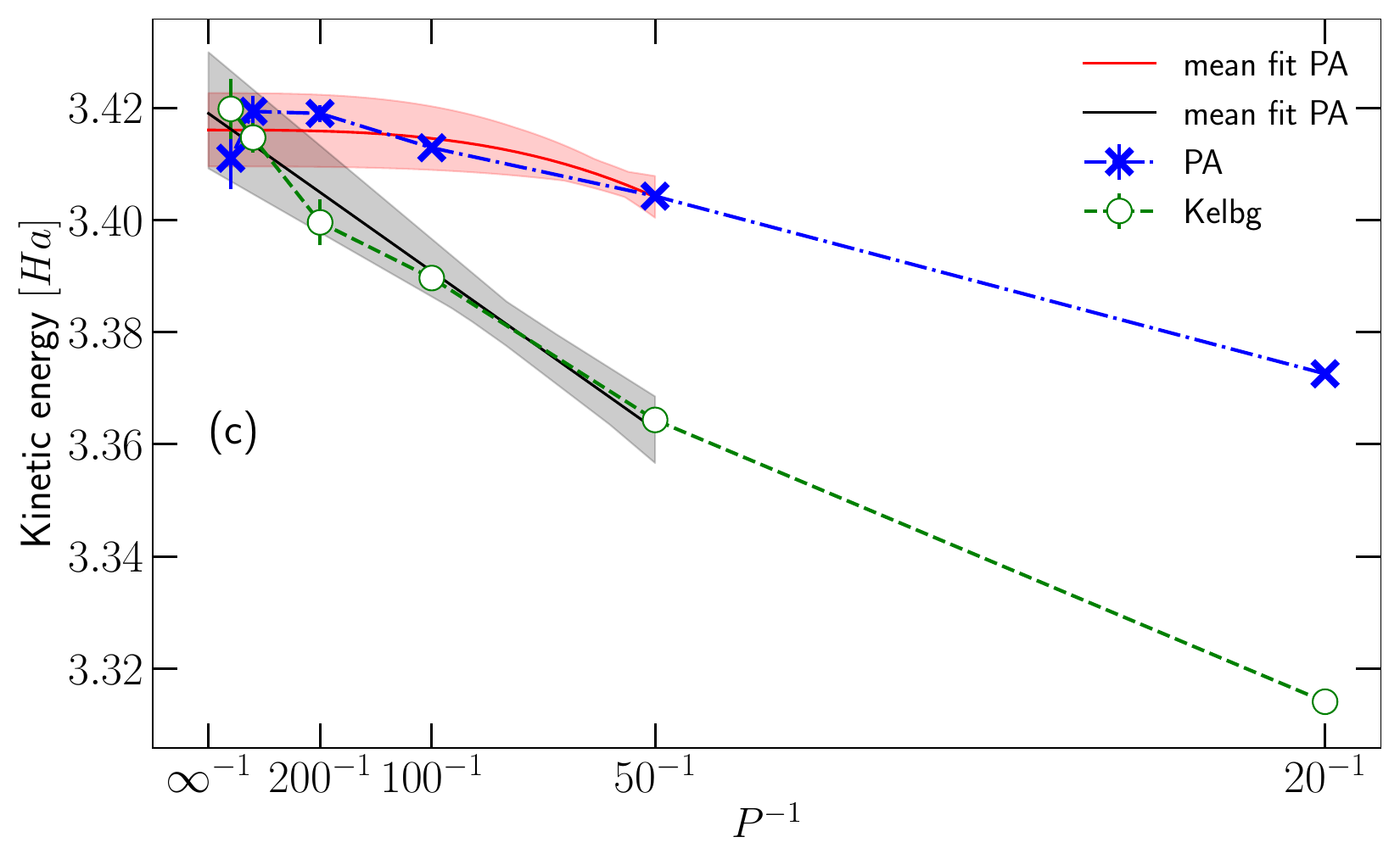}
\caption{\label{fig:convergence_plots}Convergence of (a) the total energy , (b) potential energy and (c) kinetic energy, with the inverse number of imaginary time-slices $P^{-1}$. The calculations have been carried out for $N=4$, $\Theta = 1$ and $r_s=2$. The convergence behaviour of the pair approximation is clearly favourable compared to the simple Kelbg pair potential.}
\end{figure*}

The convergence of snap-PIMC results for the energy as a function of the inverse number of imaginary time slices $P^{-1}$ is displayed in \Cref{fig:convergence_plots}, 
with the blue and green curves corresponding the PA and Kelbg action, respectively.
Figure (a) clearly demonstrates the influence of the off-diagonal contributions to the action on the convergence of PIMC expectation values. In order to account for the finite Monte Carlo errors, we carried out an extensive analysis 
of both the Kelbg potential and the pair approximation. Empirically, we find that the convergence behaviour of the Kelbg potential is proportional to $\mathcal{O}(\epsilon)$ in the given range of $P$, whereas the PA converges with $\mathcal{O}(\epsilon^3)$ which is consistent with Refs.~\cite{cep,sakkos_JCP_2009}. In order to obtain the given red and black error margins, we added Gaussian noise to the data points with the standard deviation given by the Monte Carlo error and fitted the perturbed data points to the respective convergence behaviour. This was done $10^4$ times to have a sufficient sample size. The fits were carried out starting at $P=50$ up to $P=1000$. From the obtained set of fitting coefficients, we then constructed a more fine grained $P^{-1}$ grid in the interval $I = [10^{-6}, 1/50]$. For each point in $I$, we then calculated the mean values and obtained the error margins by taking the maximum/minimum value of the obtained fit samples. This analysis has been carried out for all subplots in \Cref{fig:convergence_plots} and ensures that the errors are not underestimated. The data has been calculated up to a Monte Carlo error of a few 
millihartree.   

We already stated that the \emph{PA} includes off-diagonal terms, which are particularly important at lower temperatures. This is in contrast to the Kelbg pair-potential, which completely neglects these off-diagonal contributions. The latter only decay at large temperatures and, therefore, a large number of time slices in the PIMC simulation are needed to reach this limit. This directly explains the much faster convergence of the \emph{PA} compared to the Kelbg pair-potential in \cref{fig:convergence_plots}. While the energy estimator in the \emph{PA} is already sufficiently converged at $P = 50$ time-slice, the Kelbg estimator only shows a sufficient energy agreement with the pair approximation $P=1000$ inside the error margins. Therefore, the \emph{PA} offers a significant performance boost in comparison to the Kelbg potential, which is in qualitative agreement to earlier findings by Filinov \emph{et al.}~\cite{Filinov_PRE_2004}.
Practically, it is very hard to approach the limit of $P\to\infty$ computationally and would require at least $P=5000$ propagators, which is not feasible. 
Since the total energy is a sum of positive and negative energy contributions, see \cref{fig:convergence_plots} (b) and (c), the resulting cancellations cause a comparably large relative statistical error.
Nevertheless, if we look at the calculated potential and kinetic energy separately, we see a very good agreement within the error margins even for the very high accuracy of the data points that are of the order of a few milihartree. Our fitting extrapolation scheme indeed indicates an agreement of both the Kelbg potential and PA with increasing $P$ and furthermore confirms the faster convergence rate of the PA.

\begin{figure}
    \centering
    \includegraphics[width=0.5\textwidth]{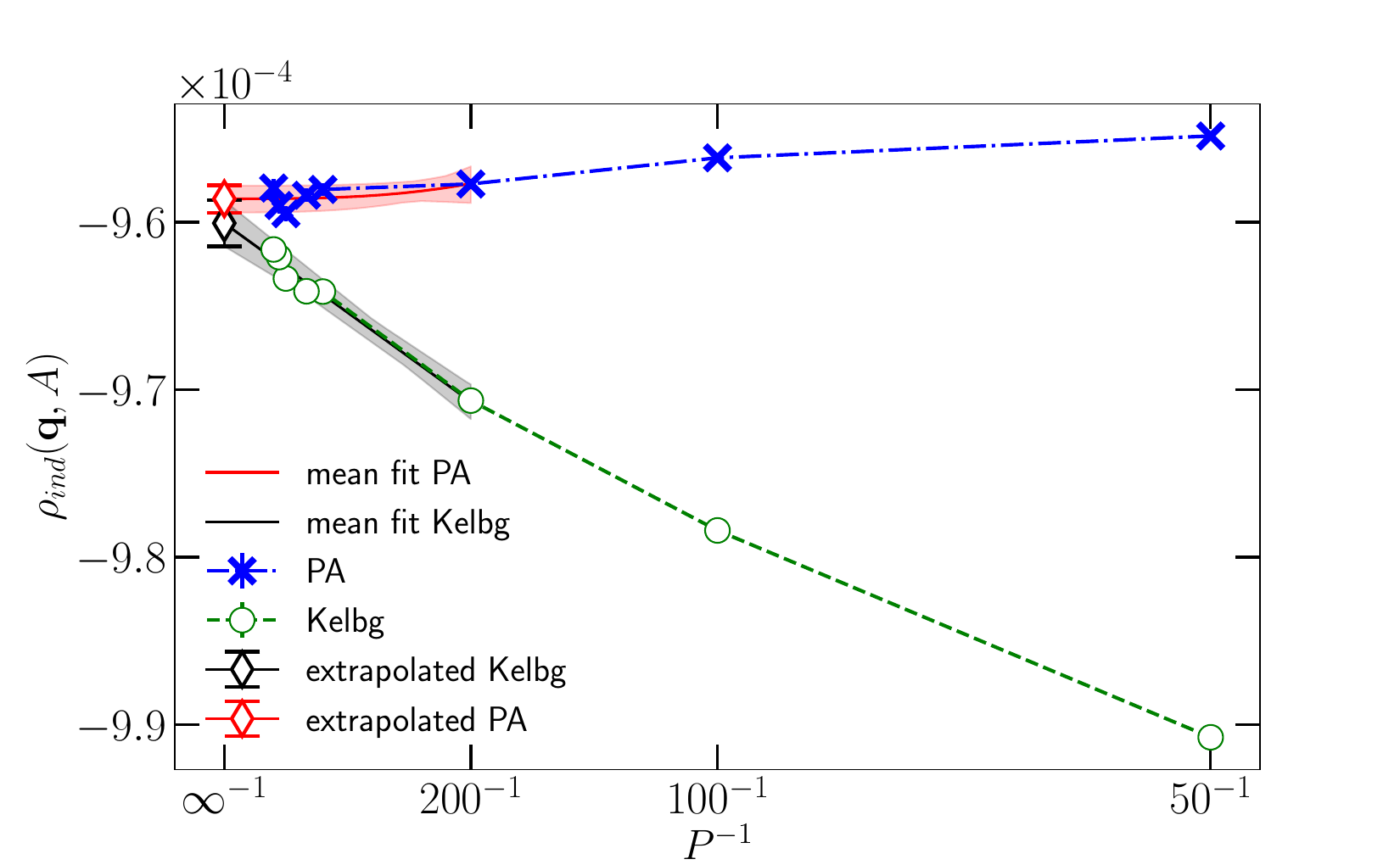}
    \caption{Induced density convergence as a function of $P$ for a perturbation strength of $A=0.1$, $N = 14$, $r_s=4$ and $\Theta = 1$. The red curve shows the results using the Kelbg pair potential and the green curve depicts the PA. Since the lower density of $r_s=4$ causes a stronger localisation of the electrons near the protons, the off-diagonal elements in the density matrix of the PA significantly improve the convergence with the number of imaginary time steps.}
    \label{fig:density_conv_induced}
\end{figure}

Another test for the convergence behaviour of both approaches is shown in \cref{fig:density_conv_induced}, where the induced density \cref{eq:rho} for $\mathbf{q} = 2\pi /$L $\hat{e}_z$ is depicted as a function of $P^{-1}$. In particular, we chose a perturbation strength of $A=0.1$ for $r_s = 4$, $N=14$ particles, and a degeneracy temperature of $\Theta=1$. The comparison between the green Kelbg curve and the blue PA curve again nicely illustrates the improved convergence rate of the PA. 
Even for the given high accuracy of the data points it is very hard to obtain an exact agreement between Kelbg and PA. For this reason, we have carried out the identical statistical analysis as described in \Cref{fig:convergence_plots}. An exact agreement with Kelbg would require a significantly higher number of propagators, which is unfortunately computationally unfeasible for the given accuracy.
Indeed, the induced density of the blue curve is well-converged already at $P=200$ within the given Monte-Carlo error bars. Due to the low density of $r_s=4$, the off-diagonal elements in the density matrix are of significant importance for the correctness of the simulation since the electrons are now more localised near the protons, which explains the substantially worse performance of the Kelbg action in this case.

\begin{figure}
    \centering
    \includegraphics[width=0.49\textwidth]{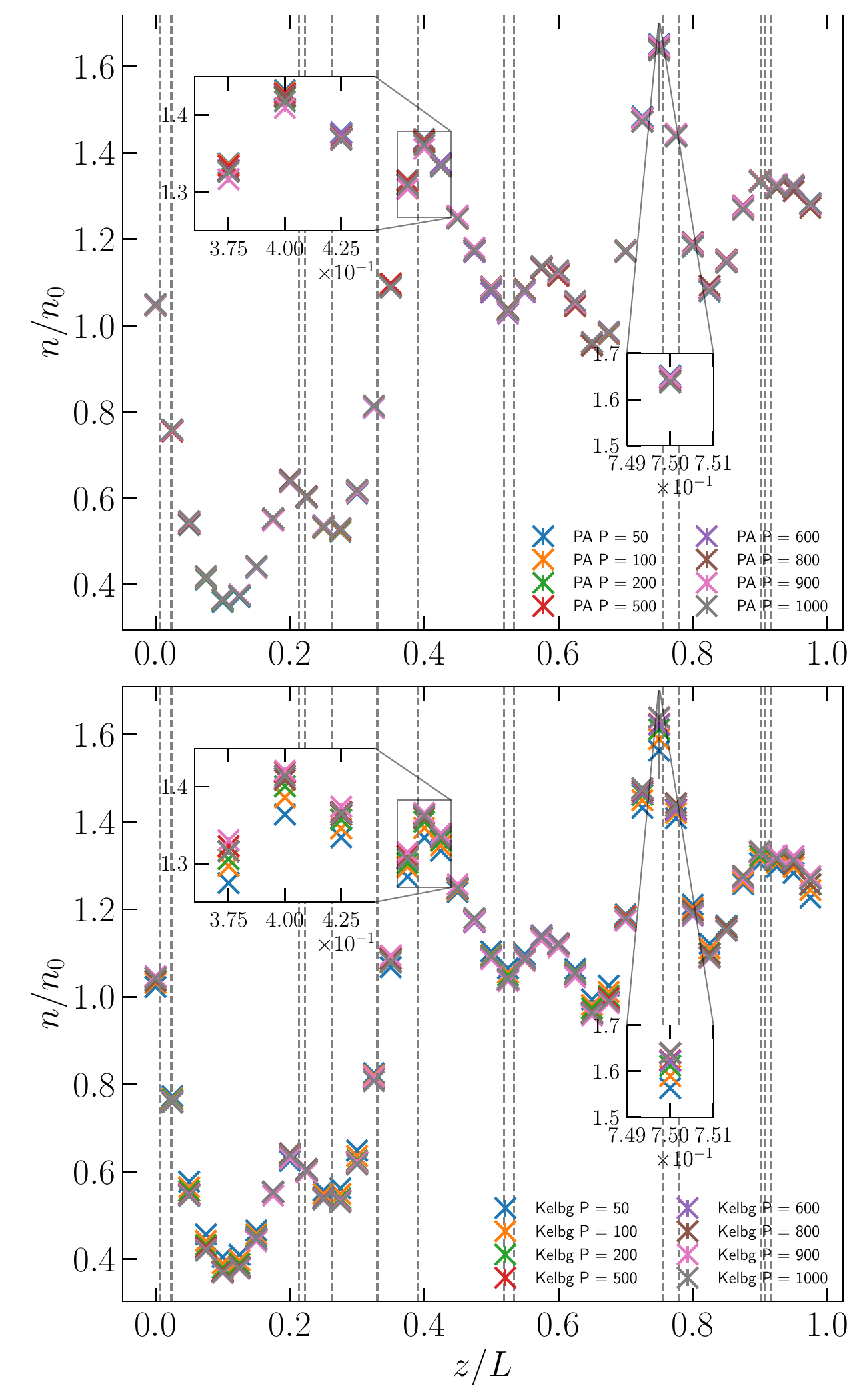}
    \caption{Density convergence with P for $r_s = 4$ and $\Theta = 1$ along the $z$-direction (i.e., integrated over $x$ and $y$). The upper panel depicts PIMC results for the density using the PA, while the lower panel shows the Kelbg density for different numbers of imaginary time-steps $P$. For both panels the insets have identical limits to highlight the improved convergence behaviour of the PA.}
    \label{fig:dens_conv_p}
\end{figure}

We now focus on the convergence behaviour of the real space density $n(\mathbf{r})$ in \cref{fig:dens_conv_p}, since it is arguably the most important observable to test the accuracy of electronic structure methods. The panels of \cref{fig:dens_conv_p} depict the electronic density (integrated over $x$ and $y$) in $z$-direction for $N=14$, $r_s=4$ and $\Theta=1$ in units of the corresponding UEG density $n_0$. The results for the density in both panels are plotted for different choices for the number of imaginary time slices in order to observe the convergence behaviour for the Kelbg pair-potential and the PA. The influence of the ions, which are shown as the vertical grey dashed lines, on the electronic localisation can clearly be observed in both panels. In the upper panel, the results for the PA are shown. We observe a very favourable convergence behaviour even for a small number of imaginary time slices. To better resolve the difference between the measured points, two insets are used to offer a better resolution around four manually chosen points. The insets reveal that all the results are already in very good agreement within the given error bars. In comparison, the lower panel shows a less favourable convergence behaviour for the density with the use of the Kelbg potential. Both inset plots in the lower panel zoom in on the identical area as above. Here we clearly observe a much more pronounced difference between the choice of imaginary time steps. We note that the error bars are an order of magnitude smaller than the estimator values and therefore are not visible in the plot. The insets were specifically chosen to be close to an ionic position in order to observe the influence of the off-diagonal contributions to the action. Since this off-diagonal error is especially pronounced in the vicinity of the ions, one can easily see that the Kelbg pair potential requires a higher number of imaginary time steps compared to the PA for a sufficient convergence. 

\begin{figure}
    \centering
    \includegraphics[width=0.485\textwidth]{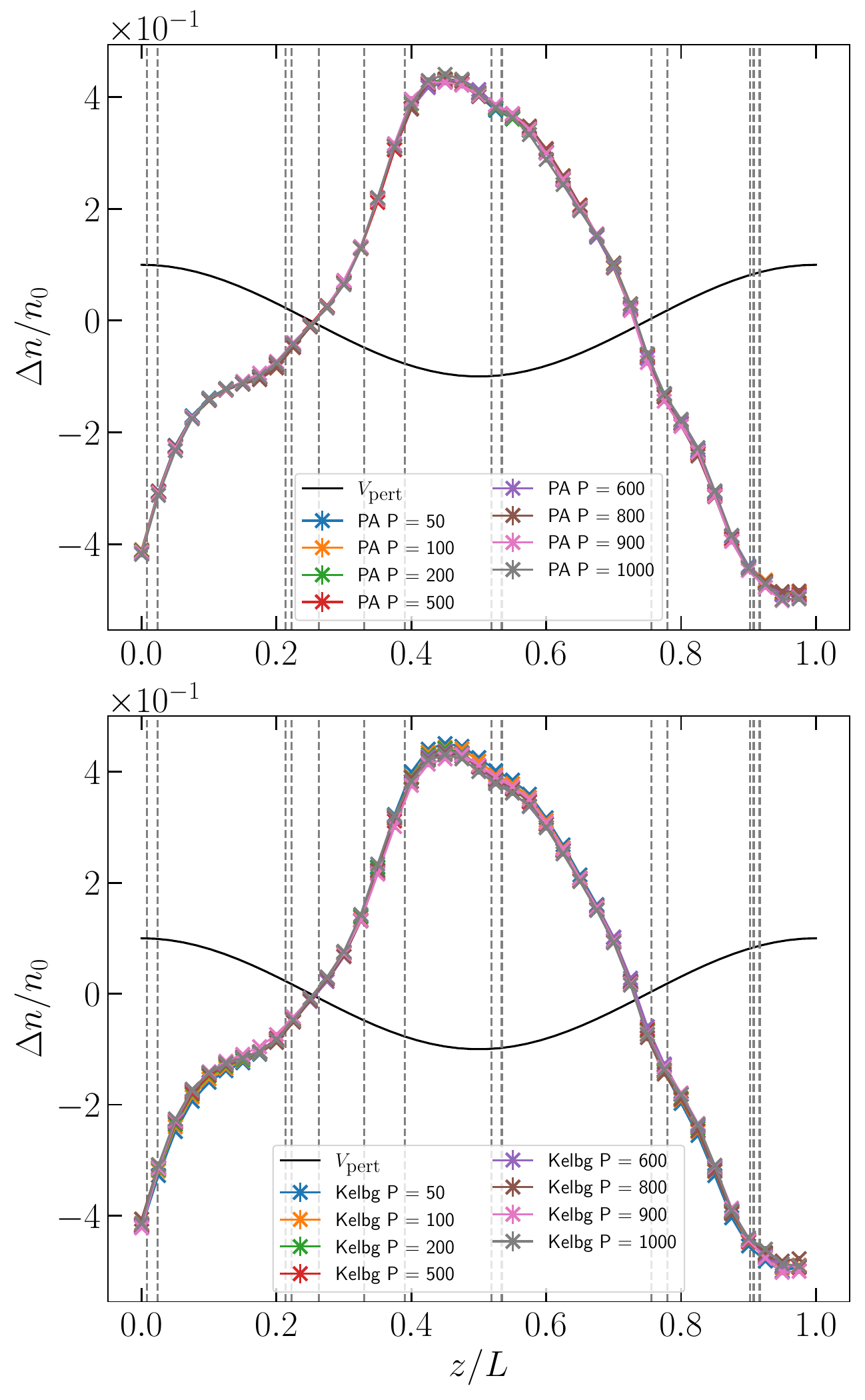}
    \caption{Change in the integrated density in z-direction due to an external perturbation of $A=0.1 \,$Ha for different $P$ at $N=14$, $\Theta=1$ and $r_s=4$. The deviation is defined as $\Delta n = n_{\textnormal{pert.}}(P) - n_{\textnormal{unpert.}}(P)$ in units of the unperturbed UEG density $n_0$. 
    In the upper panel the case of the PA is shown, while in the lower panel we depict the Kelbg result for different propagator numbers $P$. The vertical grey dashed lines show the ionic positions.}
    \label{fig:density_dev_conv}
\end{figure}

In \cref{fig:density_dev_conv}, we examine the effects of a perturbation on the electronic density. For this purpose, a perturbation of $A=0.1 \, Ha$ was introduced, while using the identical parameters as in \cref{fig:dens_conv_p}. The panels illustrate the difference in density between the perturbed and unperturbed system for the same number of imaginary time steps as in the previous figure. We define this difference in units of the unperturbed UEG density $n_0$ as 
\begin{equation}
    \Delta n(P) = \frac{n_{\textnormal{pert.}}(P) - n_{\textnormal{unpert.}}(P)}{n_0}.
\end{equation}
Again, here the grey dashed vertical lines depict the position of the ionic snapshot, and the upper (lower) panel shows the results in the case of the PA (Kelbg). The depicted quantity allows us to resolve the influence of the ions onto the response to a perturbation (black line), and the interplay with the propagator error. 
The main source of propagator error is given by the approximate treatment of the contribution of the electron-proton attraction to the total action. 
This error is proportional to the commutators of the potential with the kinetic terms in the Hamiltonian and thus proportional to the corresponding gradients. Since the impact of the external perturbation is comparably weak, these errors nearly cancel between the perturbed and unperturbed calculations, and the total propagator in $n(P)$ is much smaller.

\subsection{Fermion sign problem}
\label{ssec:FSP_results}

\begin{figure}
    \centering
    \includegraphics[width=0.48\textwidth]{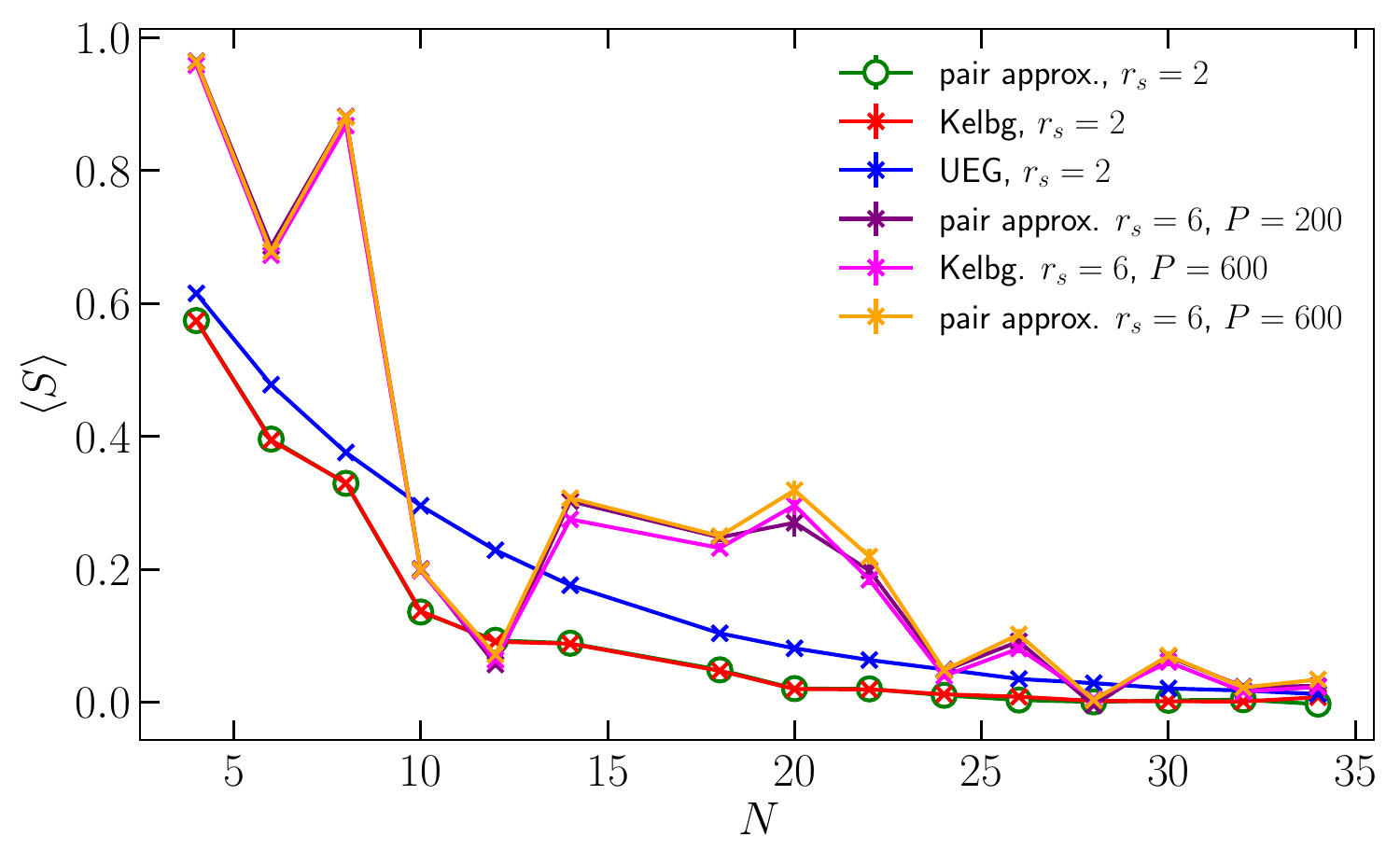}
    \caption{The average sign $\langle S \rangle$ compared both with the Kelbg potential (red crosses) as well as the \emph{PA} (green circles) at metallic densities $r_s = 2$ and $\Theta = 1$. Both curves exhibit an exponential decay of the average sign with increasing particle number. The Kelbg result has been calculated using $P = 600$ propagators, where the \emph{PA} only required $P=200$ propagators. Furthermore, the same conditions have been studied for $r_s=6$. The magenta crosses shows the Kelbg result at $r_s=6$ again with 600 propagators. For the PA we ran the same calculations but for $P=200$ (purple crosses) and $P=600$ (yellow crosses).}
    \label{fig:sign_N_theta_1}
\end{figure}

\begin{figure}
    \centering
    \includegraphics[width=0.485\textwidth]{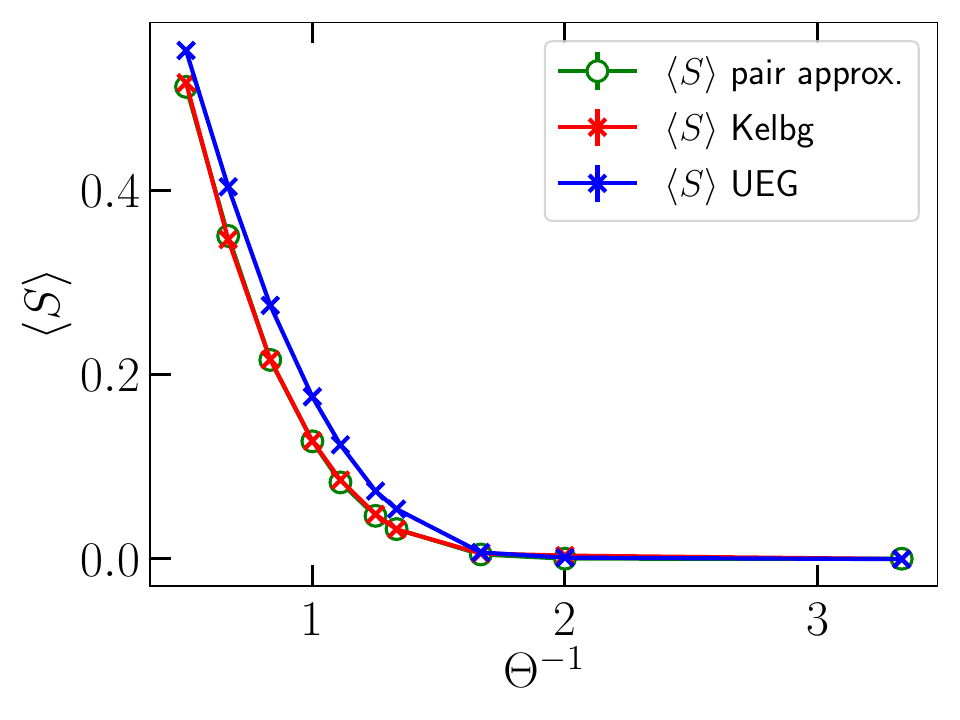}
    \caption{Average sign $\langle S \rangle$ as a function of the degeneracy parameter $\Theta=1$ for $N = 14$ particles at metallic density $r_s = 2$. The Kelbg (red) and the \emph{PA} (green) results show the expected increase of the average sign. The \emph{PA} used $P=200$ and the Kelbg calculations were run with $P=600$ propagators.}
    \label{fig:sign_theta_14}
\end{figure}

The fermion sign problem constitutes the main computational bottleneck of our simulations.
In \cref{fig:sign_N_theta_1}, we show the 
dependence of $\braket{\hat S}$ on the particle number 
for a temperature of $\Theta=1$ and the metallic density $r_s=2$. 
First and foremost, we note that all curves exhibit a qualitatively similar exponential decay with $N$, cf.~Eq.~(\ref{eq:sign_avg}).
Moreover, both the average sign of the Kelbg implementation as well as the \emph{PA} are in excellent agreement with each other. 
A similar result can be seen in \cref{fig:sign_theta_14} 
where the sign exponentially decays with the inverse temperature $\Theta^{-1}$.
Since the value of the average sign is very sensitive to the sampling of the permutation space~\cite{Dornheim_permutation_cycles}, the excellent agreement shows that both the \emph{PA} and the Kelbg implementation are sampling the canonical partition function $Z$ nearly identically. The only difference here is, that the number of propagators required for the \emph{PA} ($P=200$) is much lower compared to the diagonal Kelbg potential ($P=600$). Thus, the \emph{PA} has a distinct performance advantage over the Kelbg pair-potential. 

Let us next consider the blue lines in \cref{fig:sign_theta_14,fig:sign_N_theta_1}, which  are depicting the average sign of the corresponding uniform electron gas simulations at the same density, $r_s=2$. 

The sign of the UEG simulations in \cref{fig:sign_N_theta_1} is consistently increased compared to the snapshot results. This is caused by the decrease in uniformity of the hydrogen system, since the presence of protons causes an increased localisation of the electronic imaginary time paths around them. 
In particular, the probability to form a permutation cycle is proportional to the distance between two beads of different particles.
For the case of the UEG, the paths are uniformly distributed and, therefore, the average sign follows the exponential decay proportional to the particle number, see \cref{eq:sign_avg}. Nonetheless, the protons positions in the snap-PIMC simulations induce a non-uniformity in the electronic imaginary time paths. At $r_s=2$ the somewhat increased localisation of electronic paths therefore increases the acceptance probability of an exchange cycle.
We want to further study how localisation around the protons affects the average sign. For this reason the purple curve in \cref{fig:sign_N_theta_1} illustrates the average sign of the hydrogen snapshot at $r_s = 6$ using $P=600$ propagators with $N=14$ particles at $\Theta = 1$. As a comparison, the magenta and yellow line show results for the same conditions at $P=600$ using Kelbg and PA,  respectively. All three curves are in good agreement with each other. We can observe in all three lines a severe snapshot dependency, since at the low density of $r_s = 6$ a significant increase of the localisation around the protons is occurring. If two protons are close to each other, then the electronic paths around the ions will be closer together as well and therefore increase the probability of an exchange cycle being sampled. However, in case the protons are farther away from each other, the probability for an exchange cycle being accepted decreases. Therefore, one observes an oscillation around the exponential decay of the sign at $r_s=6$ due to the snapshot dependency. We strongly suspect that the inclusion of dynamical ions in PIMC would possibly lead to a more severe sign problem associated with the formation of molecules.

\subsection{Impact of the density parameter $r_s$}
\label{ssec:rs}
\begin{figure}\centering
\includegraphics[width=0.485\textwidth]{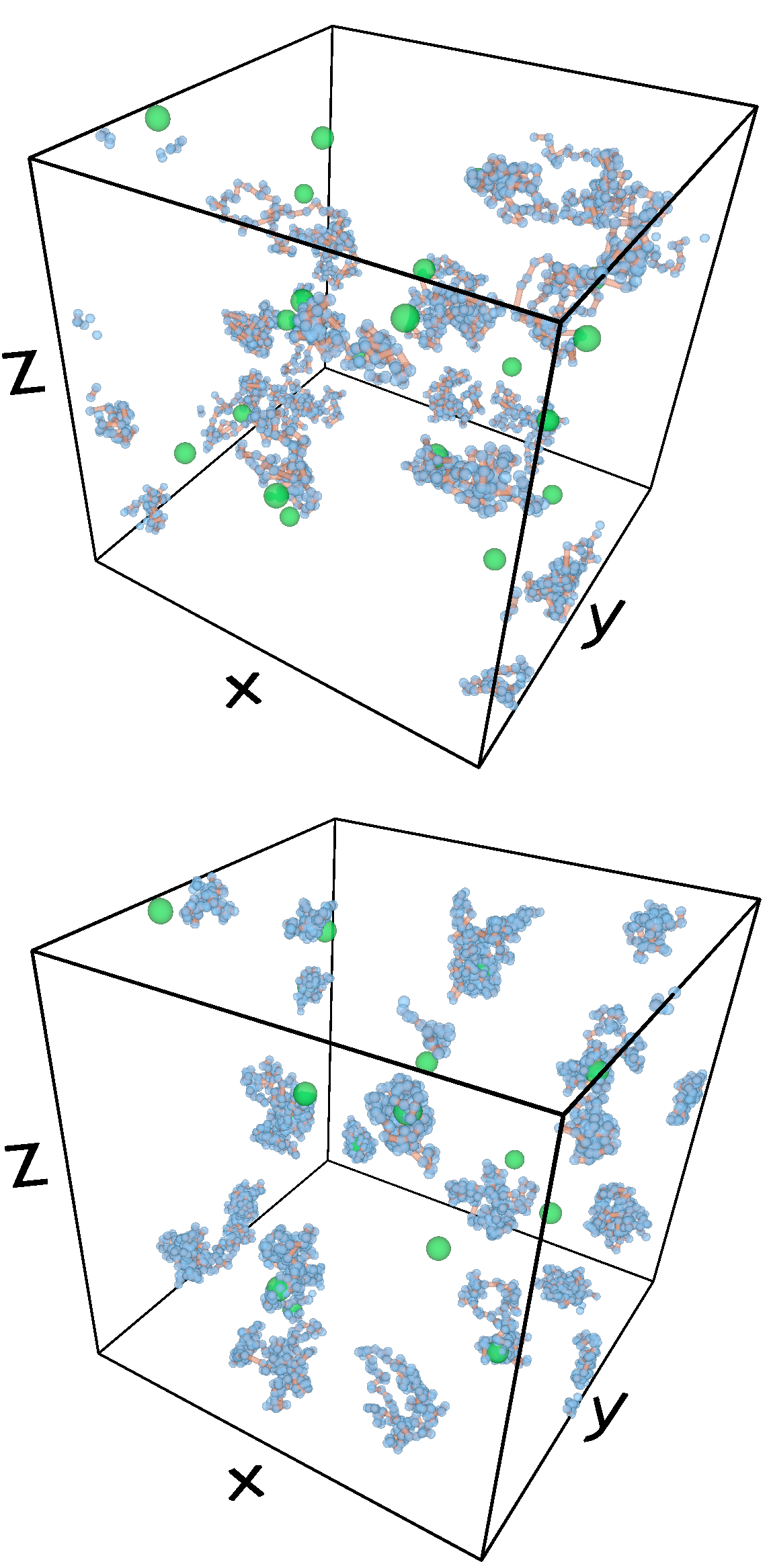}
\caption{\label{fig:snaps}
Configurations from pair approximation PIMC simulations of a hydrogen snapshot with $N=20$, $\theta=1$, and $P=200$ for $r_s=2$ (top) and $r_s=6$ (bottom). The green spheres denote the ion positions, and the blue paths depict the imaginary-time paths of the electrons. At $r_s=6$, the localisation of the electron paths around the ions is substantially increased compared to $r_s=2$. This heuristically indicates the emergence of \emph{bound states} around the ions at lower density. The top panel has been taken from the Supplemental Material of Ref.~\cite{Boehme2022}.
}
\end{figure}

In order to show how the density parameter influences the imaginary-time paths in the PIMC simulations, we show in \cref{fig:snaps} a hydrogen snapshot at $r_s=2$ (top) and $r_s=6$ (bottom) for $N=20$, $P = 200$ and $\Theta = 1$. The imaginary-time paths are useful to gain a qualitative understanding of the electronic behaviour of the system. The top panel at $r_s=2$ shows more disordered paths compared to the bottom panel, where the electronic paths are substantially localised around some of the ions. This result shows heuristically the emergence of \emph{bound states} for lower density and sufficiently low temperatures, since the path localisation gives a measure for the estimated electronic density. Nevertheless, we stress that PIMC does not make the artificial distinction between bound and free states.

\begin{figure}\centering
\includegraphics[width=0.485\textwidth]{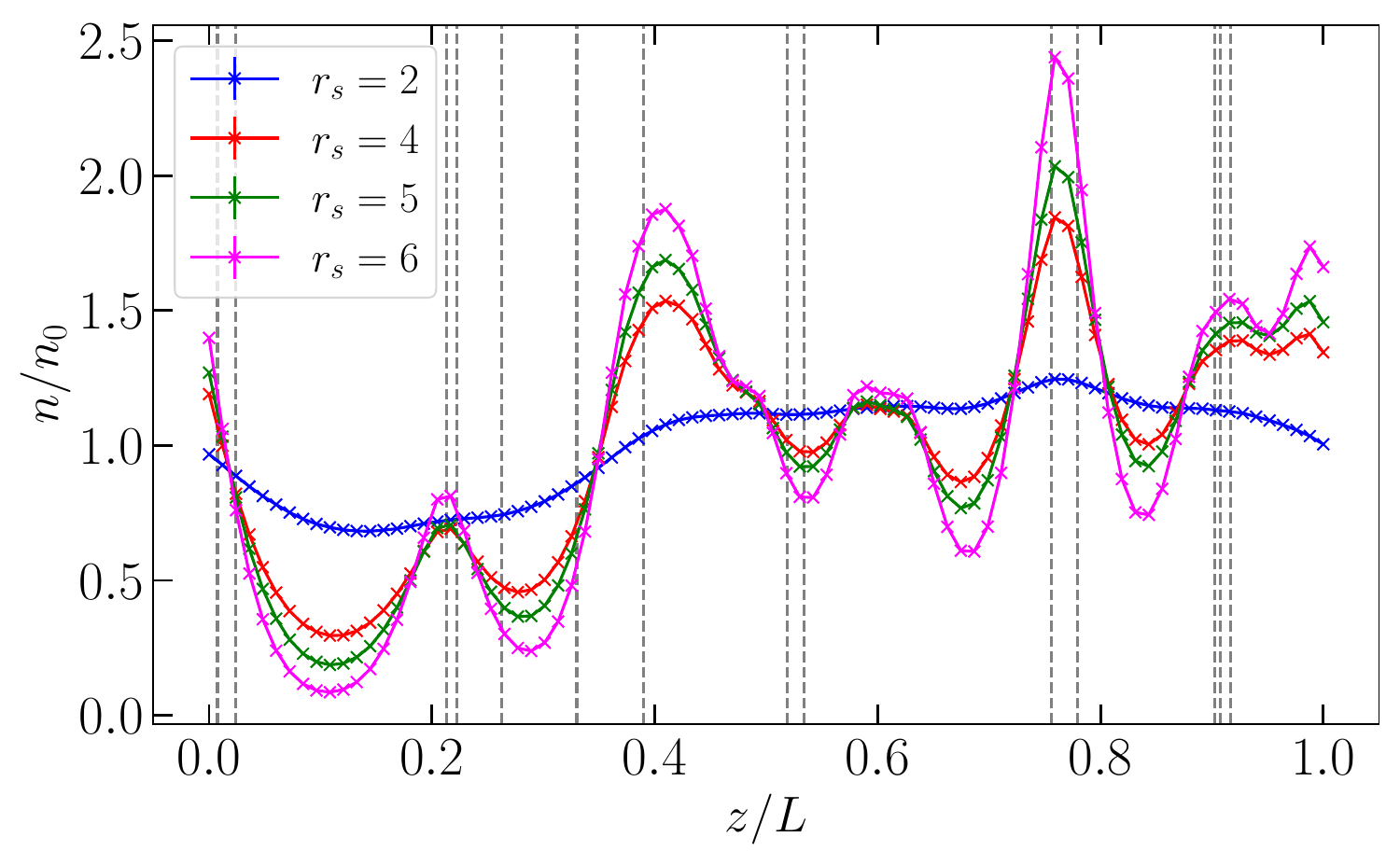}
\caption{\label{fig:density_comparison_rs}
Comparing PIMC results for the density along the $z$-direction for the same fixed ion snapshot with $N=14$ at $\theta=1$ for different values of the density parameter $r_s$. The dashed vertical lines show the $z$-coordinates of the ions. 
}
\end{figure}

A more quantitative picture of the density induced increased localisation around the ions is depicted in \cref{fig:density_comparison_rs} for $N=14$ and identical parameters as in the previous figure. The figure visualises the electronic density for a hydrogen snapshot along the z-axis in units of the unperturbed UEG density. For this run, the \emph{PA} was used using $P=600$ propagators. The figure clearly indicates an increased electronic density around the ion positions with an increasing value for $r_s$. We especially find the strong localisation difference at $r_s=2$ compared to $r_s=4$. As already stated for \cref{fig:snaps}, this illustrates how electrons start to localise around the protons for decreasing densities, which would result in \emph{bound states} in a simplified single-particle picture.

\subsection{Impact of the temperature parameter $\Theta$}
\label{ssec:theta}
\begin{figure}\centering
\includegraphics[width=0.485\textwidth]{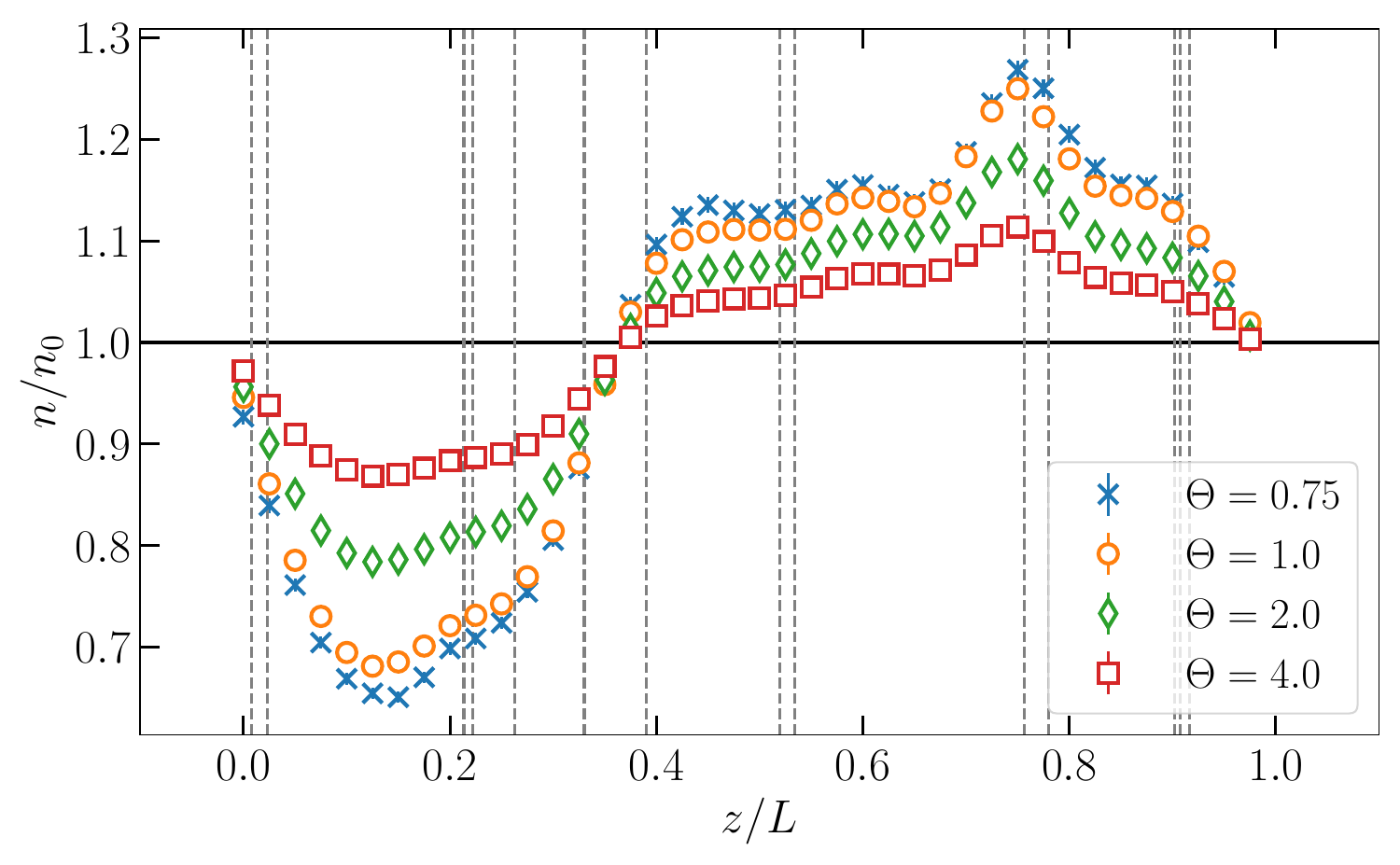}
\caption{\label{fig:density_comparison_rs2_theta}
Comparing PIMC results for the density along the $z$-direction for an ion snapshot of $N=14$ at $r_s=2$ for different values of the reduced temperature $\theta$. The dashed vertical lines show the $z$-coordinates of the ions.
}
\end{figure}

The impact of temperature on the electronic density along the $z$-coordinate is shown in \cref{fig:density_comparison_rs2_theta}  for a snapshot with $N=14$, $r_s=2$  and different $\Theta$ values. An increase in temperature clearly leads to the trend of the snapshot density converging towards the UEG density. This can be explained by the fact that with increasing temperature the influence of the ions becomes the less dominant, since thermal excitations are getting stronger than the electron-ion interaction.

\subsection{Comparison to DFT}
\label{ssec:DFT}

\begin{figure}\centering
\includegraphics[width=0.485\textwidth]{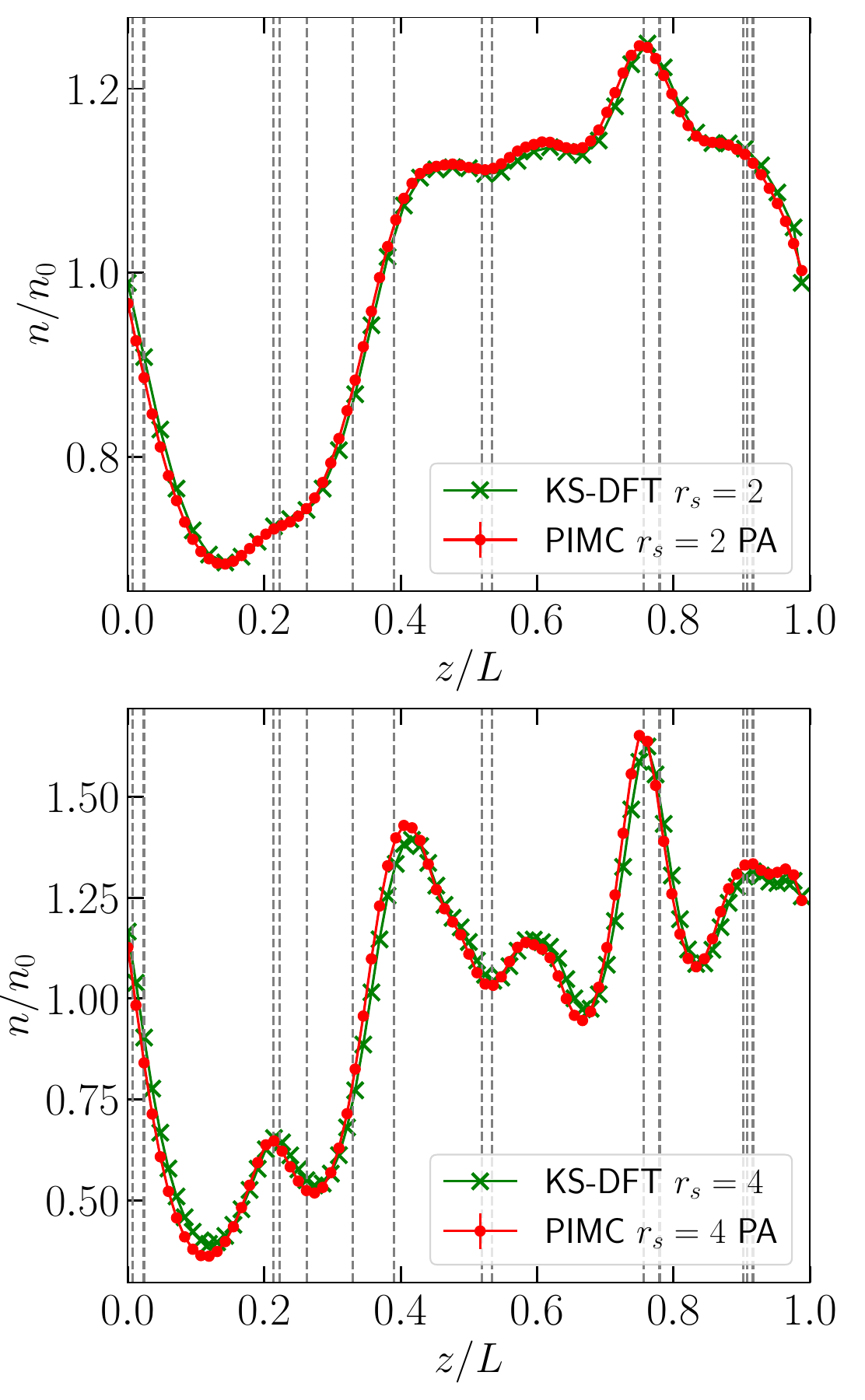}
\caption{ \label{fig:density_strip_N14}
Top (bottom): comparison of the density along the $z$-direction of an unperturbed snapshot with $N=14$ unpolarized electrons at $r_s=2$ ($r_s=4$) and $\theta=1$ between thermal DFT within LDA (green crosses) and PIMC with the PA and $P=600$ (red circles).
}
\end{figure}

Our exact PIMC solutions to the many-electron problem in the external ion snapshot potential put us into the unique position to benchmark the accuracy of thermal DFT calculations for exactly the same Hamiltonian. This is shown in \cref{fig:density_strip_N14} for the density (integrated over $x$ and $y$) along the $z$-direction; we re-iterate our earlier point that the electronic density constitutes the central observable within DFT, and its importance can hardly be overstated.
In the upper panel, the red dots show 
our PIMC results obtained within the PA in units of the unperturbed UEG density at $r_s=2$. The green line shows the resulting real-space density from DFT using LDA. Evidently, both results are in excellent agreement with each other. This somewhat changes at $r_s=4$ depicted in the lower panel. In the vicinity of protons (grey dashed lines), one can observe a small yet significant disagreement between snap-PIMC and DFT. The reason for this is, that the employed LDA functional is based on the ground-state UEG data from Ceperley and Alder \cite{CeperleyAlder}. Therefore, it is not possible for LDA to fully capture the impact of the inhomogeneity around the protons.
The KS-DFT overestimates the spreading degree of the electronic density around  protons due to the known delocalization (self-interaction) error  of commonly used XC approximations in KS-DFT \cite{doi:10.1126/science.1158722}.  This failure is particularly stark for the calculation with lower densities as it is demonstrated for $r_s=6$ by B\"ohme \textit{et al.} \cite{Boehme2022} (see Supplement).
This demonstrates the importance of rigorous benchmarks of commonly used XC functionals in the WDM regime, in particular for lower densities. Recently, a more detailed analysis of the performance of various LDA, GGA, and meta-GGA level XC functionals in the case of hydrogen in terms of the density response function was reported by Moldabekov \textit{et al.}~\cite{DFT_kernel_22}. 
 For $r_s=2$, since it is a metallic density, the electrons behave qualitatively similar to a UEG~\cite{Boehme2022}, and the DFT results are in much better agreement with the snap-PIMC density.

\section{Summary and Discussion, Outlook}
\label{sec:summary}
In this work, we gave a comprehensive introduction to 
\emph{ab initio} PIMC simulations of hydrogen snapshots in the WDM regime. 
To avoid the notorious path collapse due to the Coulomb attraction between electrons and protons, we employ the pair approximation, which exhibits a favourable convergence behaviour compared to the simpler Kelbg potential.
Furthermore, we do not impose any nodal restrictions on the paths, which means that our simulations are afflicted with the fermion sign problem. Still, simulations are feasible over a substantial part of the relevant parameter space.

From a physical perspective, we find an increased electronic density around the proton positions for $r_s=4$ compared to the metallic density $r_s=2$. 
A similar effect has been observed for decreasing temperature.
In addition, we have compared our exact new snap-PIMC results for the electronic density to thermal DFT calculations for the same snapshot 
and found excellent agreement at $r_s=2$. However, the electronic density using LDA  significantly deviates from our results especially in the vicinity of the protons at $r_s=4$. The reason for this can be traced back to the fact that LDA stems from a UEG ground state calculations and, by definition, cannot fully account for ionic influences. We note that our PIMC results constitute an ideal benchmark for the assessment of different XC-functionals~\cite{Moldabekov_JCP_2021,Moldabekov_PRB_2022}, which will be pursued in more detail in future works.


A key advantage of the snap-PIMC approach is the straightforward access to the exact static electronic response of hydrogen, as we have demonstrated in Ref.~\cite{Boehme2022}. This opens up the enticing opportunity to compute the exact XC-kernel of a real material in the WDM regime, which in turn can be utilised in a number of applications such as \textit{linear response time dependent DFT} simulations. The latter, in turn, give one access to the dynamic structure factor, which is the central property in modern XRTS experiments~\cite{siegfried_review,kraus_xrts}.

Snap-PIMC can easily be extended beyond hydrogen, as the computation of the two-body density matrix is possible for all elements. However, one needs to keep in mind that the Ewald sum requires charge-neutrality and therefore the required amount of electrons increases drastically. The fermion sign problem then might make simulations computationally unfeasible in the case of too heavy elements. 
Nevertheless, we note that exact PIMC simulations of deuterium--tritium mixtures and helium constitute a realistic prospect.
We will further investigate the possible extension of this approach to a full two-component PIMC component, which would enable us to capture the exact static response quantities including ionic contributions.

Eventually, we will extend the current set-up two a full two-component PIMC simulation of hydrogen, where the ions are treated on the same footing as the electrons, i.e., are not fixed.
For example, this will allow us to compare to widely used DFT-MD simulations with respect to different properties such as energies and pressure.
Additionally, this implementation will enable the extraction of exact screened potentials~\cite{doi:10.1063/5.0097768}, which are a highly relevant object for the quantum statistics of plasmas.


\section*{Acknowledgments}
This work was partly funded by the Center for Advanced Systems Understanding (CASUS) which is financed by Germany's Federal Ministry of Education and Research (BMBF) and by the Saxon Ministry for Science, Culture and Tourism (SMWK) with tax funds on the basis of the budget approved by the Saxon State Parliament.
The PIMC calculations were carried out at the Norddeutscher Verbund f\"ur Hoch- und H\"ochstleistungsrechnen (HLRN) under grant shp00026 and on a Bull Cluster at the Center for Information Services and High Performance Computing (ZIH) at Technische Universit\"at Dresden.

\bibliography{bibliography}

\end{document}